\documentclass[aps,pre,amsmath,amsfonts,amssymb,floatfix,superscriptaddress,nofootinbib]{revtex4}
\usepackage{mathrsfs} \usepackage{graphicx,color} \usepackage{bbm} \usepackage{multirow}
\usepackage{bbold}

\let\a=\alpha \let\b=\beta \let\g=\gamma \let\d=\delta
 \let\z=\zeta \let\h=\eta \let\k=\kappa
\let\l=\lambda    
\let\s=\sigma  \let\f=\varphi 
   \let\G=\Gamma
\let\D=\Delta \let\Th=\Theta  
   
 \let\r=\rho \let\th=\theta

\def\FF{{\cal F}}

\def\DD{{\cal D}} \def\SS{{\cal S}}

  \def\erf{\text{erf}}

\def\de{\mathrm d}

\def\eee{\mathrm{e}}

\def\to{\rightarrow} \def\la{\left\langle} \def\ra{\right\rangle}

\newcommand{\beq}{\begin{equation}} \newcommand{\eeq}{\end{equation}}
\newcommand{\wh}{\widehat}

\newcommand{\scl}{\D_{EA}^{\frac{1}{\kappa}}}

\allowdisplaybreaks

\begin{document}

\title{
Following the evolution of glassy states under external perturbations: \\the full replica symmetry breaking solution
}

\begin{abstract}
The state-following technique allows the study of metastable glassy states under external perturbations. Here we show how this construction can be used to study the behavior of glassy states of Hard Spheres in infinite dimensions under compression or shear strain. 
In \cite{RUYZ15} it has been shown that in both cases, when the external perturbation is sufficiently strong, {glassy states undergo a second-order transition, called the Gardner transition, whereupon a hierarchical structure of marginal micro-states manifests within the original glass state}. The purpose of this work is to study the solution of the state-following construction in this marginal phase. {We show that upon} compression, close to the jamming transition, the metastable states are described by a scaling solution characterized by a set of non-trivial critical exponents that agree with the results of \cite{CKPUZ14NatComms}, {and we compute the value of the jamming density $\wh \f_j$ for various glassy states.} Moreover we show that under the action of the shear strain, beyond the Gardner point, the metastable states can be followed in the marginal phase {and we detect an overshoot in the stress-strain curve in agreement with numerical and experimental observations}. Finally we further characterize the Gardner transition point by computing both the $\chi_4$ susceptibility and the exponent parameter $\lambda$ {that characterize the critical slowing down of the dynamics within a glassy state close to the transition}.

\end{abstract}

 \author{Corrado Rainone}
 \affiliation{Department of Physical Chemistry, the Weizmann Institute of Science, Rehovot 76100, Israel}

\author{Pierfrancesco Urbani}
\affiliation{Institut de Physique Th\'eorique, CEA, CNRS, F-91191 Gif-sur-Yvette Cedex, France}

%
%
%

\maketitle

\section{Introduction}
The huge dynamical slowing down that is observed in supercooled liquids undergoing the glass transition can be explained in terms of the appearance of a large number of metastable states \cite{KT87,KT87b,KT88,KTW89}, {which} deeply affect the structural relaxation of the system.
When a supercooled liquid is close to the glass transition point its dynamics can be thought as an initial fast relaxation in one of the closest typical metastable states available, followed by a very slow {hopping} from one metastable state to the other.
Since metastable states are long lived, {this induces} a separation of time scales that allows the introduction of quasi-equilibrium techniques to study the properties of glassy states of matter.\\
Practically, one can think of a glass as an amorphous structure. The particles can thus only vibrate around the {nodes} of {an} amorphous lattice, and the structural relaxation is \emph{effectively} frozen since it happens on much longer timescales with respect to {these} local vibrations.
This means that if we neglect the activated jumps from one metastable state to the other, we can describe glasses by computing the typical properties of the phase space associated to the vibrations of the particles around a typical amorphous structure. This can be done {in practice employing} the so-called Franz-Parisi construction.

This state following procedure has been first developed in \cite{FP95} and very recently it has been discussed in the case of {diluted} spin glass models in \cite{KZ10,KZ10b}. The {basic} idea is to consider a \emph{master} system, whose configurations are extracted with an equilibrium measure, and a \emph{slave} system whose configurations are again equilibrated, but constrained to be close to the ones of the master system. This way the master system {plays the role of} the amorphous lattice and the slave one probes the available phase space of vibrations around it. This construction has been firstly developed in the context of schematic spin-glass models \cite{FP95,BFP97,KZ10,KZ10b,KZ13} and then it has been progressively applied to structural glasses \cite{CFP98} using specific approximation schemes of static liquid theory. Moreover it has been used to describe the long time regime of glassy dynamics \cite{FP13,FPU15}.

In a very recent work \cite{RUYZ15}, we have used this formalism to study how glassy states behave under {adiabatic} external perturbations.
We have considered a prototypical model of structural glasses, namely a system of hard spheres, and we have studied the behavior of glassy states both under compression and upon application of a shear strain.
The results of this analysis have shown that for sufficiently strong external perturbations, each glassy state undergoes a phase transition, referred to as the \emph{Gardner transition} \cite{Ga85, GKS85}. This is a critical point whereupon the phase space that belongs to a metastable glassy state gets fractured into a hierarchical landscape of substates. {The presence of this transition had been already detected} in {mean-field} hard spheres in \cite{KPUZ13}, and in \cite{CKPUZ14NatComms, CKPUZ14JSTAT} {it has been shown} how this transition deeply affects the critical properties of hard spheres close to the jamming point, {culminating} in the exact calculation, {from first principles}, of the critical exponents of the jamming transition \cite{LN01,LNSW10,LDW13}.  

In the present work we want to characterize the evolution of glassy states beyond the Gardner transition point.\\
Our starting point is again the hard sphere model in infinite dimensions. The main reasons to look for this mean field limit are two: first of all it makes the model exactly soluble, and secondly, it guarantees a precise definition of metastable states in terms of local minima of a free energy landscape \cite{SSW85}. 
This last point is the most important one. Indeed, in finite dimensions every free energy functional (obtained with a Legendre transform) must be a convex function of any order parameter. The non-convexity can survive only in a mean field limit. 
The validity of mean field theory depends crucially on how much metastable states are long-lived and, in particular, how important are strong perturbative and non-perturbative effects like activation and nucleation; from a quantitative level this can be understood by introducing a suitable Ginzburg criterion \cite{FJPUZ12}, {as per usual practice in a renormalization group setting}.\\
However, in the present work we completely neglect the issue of finite dimensional fluctuations whose systematic inclusion is a far reaching ({and very long-standing}) problem and we focus only on mean field theory that provides the starting point of any finite dimensional calculation.\\
The paper is organized as follows: in the first section we discuss the state-following procedure using the Franz-Parisi (FP) potential; then, {building up from the results of} \cite{RUYZ15}, we compute its expression beyond the Gardner transition point;
we then show the results obtained from the study of the FP potential in the Gardner phase, and finally we compute the behavior of the $\chi_4$ susceptibility and the dynamical critical exponent on approaching the Gardner transition point.

\section{The state following procedure: a general formalism}
The FP potential for a system of hard spheres has been introduced and discussed in \cite{RUYZ15}.
Here we briefly review this construction.
 
We consider a \emph{master} system of hard spheres whose positions are denoted as $R=\{r_i\}_{i=1,\ldots, N}$. Moreover we consider $m-1$ replicas of the master system. This way we have $m$ replicas whose coordinates are given by $R^{a}=\{r_i^\a\}_{i=1,\ldots, N}^{a=1,\ldots, m}$ being $R^1$ the configuration of the original master system.
The underlying reason to do so is that we can in principle use the analytic continuation of the number of replicas $m$ to non-integer values to tilt the Boltzmann measure on a particular class of metastable glassy states \cite{Mo95, KPUZ13,Za10}. {This would enable us to select a glass state point out of equilibrium, corresponding for example to a rapid quench followed by an adiabatic perturbation. However, we will be mainly concerned with the $m=1$ case wherein the initial state point is chosen at equilibrium, corresponding to a slow annealing followed by a pertubation. We leave the $m\neq 0$ case for future work}.
The spheres in each replica interact via a hard core potential $V(R^a)=\sum_{i<j}v(r_i^a-r_j^a)$ and all of them are at the same temperature $T_g$ and reduced packing fraction $\wh \varphi_g$\footnote{The reduced packing fraction is defined from the real packing fraction $\varphi$ as $\wh \varphi = 2^d \varphi /d$. In the infinite dimensional limit it can be shown that this is the good scaled control parameter \cite{PZ10}.}. Instead, two spheres that belongs to two different replicas interact through a vanishing attractive coupling that will be enough to let them fall down inside the same metastable glassy state\cite{Mo95}.\\
Then we consider a \emph{slave} system of hard spheres whose positions are denoted with $X=\{x_i\}_{i=1\ldots, N}$. 
If the master replicas are trapped inside a metastable glassy state we can use the slave system to probe the phase space that belongs to it.
A simple way to do this is to constrain the positions of the spheres in the slave system to be close to the positions of the spheres belonging to the first replica among the master ones.

The appropriate distance between the configuration of the slave and the master systems is given in terms of the mean square displacement (MSD)
\beq
\Delta(X,R^1)=\frac{d}{N}\sum_{i=1}^N|x_i-r_i^1|^2
\eeq 
where $d$ is the spatial dimension that has been included in the definition in order to have a finite quantity in the $d\to \infty$ limit \cite{KPZ12}.

The slave system can be at a different temperature $T_s$, different reduced packing fraction $\wh \varphi_s=\wh \varphi_g \eee^\eta$ and it can be also subjected to an external strain $\g$. We thus denote the interaction potential of the slave system as $V_{\g,\eta}(X)=\sum_{i<j}v_{\eta,\g}(x_i-x_j)$ \cite{RUYZ15}.

In the FP scheme, we are interested in computing the free energy associated to the slave system that is constrained to be at a distance smaller than, say, $\Delta^r$ from the configuration of the master system. 
{This means that} we are interested in the following quantity
\beq
-\b_s N F_g[\eta,\g, \beta_s, \Delta^r ]=\overline{\log \int \de X \eee^{-\beta_s V_{\eta,\g}(X)}\delta[\Delta^r-\Delta(X,R^1)]}
\eeq
where the overline denotes the average over $R^1$ that is defined with the replicated measure
\beq
\begin{split}
\overline{O(R^1)}&=\frac{1}{Z_m}\int \left(\prod_{a=1}^m\de R^a\right) O(R^1) \eee^{-\b_g\sum_{a=1}^m V(R^\a)}\\ 
Z_m&=\int \left(\prod_{a=1}^m\de R^a\right) \eee^{-\b_g \sum_{a=1}^mV(R^a)}
\end{split}
\eeq

Note that $R^1$ acts as a quenched disorder on the slave system. The average over the configuration $R^1$ is made difficult by the presence of the logarithm. This difficulty can be overcome by replicating the slave system $s$ times. 
Let us define
\beq
-\b_s N F_{FP}=\log \int\left(\prod_{\a=1}^m\de R^a\right)\left(\prod_{b=1}^s\de X^s\right) \eee^{-\b_g \sum_{a=1}^m V(R^a)-\b_s\sum_{b=1}^sV_{\g,\eta}(X^b)}\left[\prod_{b=1}^s\delta\left(\Delta^r-\Delta(X^b,R^1)\right)\right]
\eeq
Then we have that
\beq
F_g= \lim_{s\to 0}\partial_s F_{FP}
\label{limit_sto0}
\eeq
as usual with the replica trick. $F_g$ is the free energy of the glassy state that has been \emph{planted} at $(\wh\f_g,T_g)$ and then \emph{followed} in compression and/or shear strain. 
Thus, {our problem is now to compute} the free energy of a system of $m+s$ replicas of a hard spheres.

\section{Computation of the entropy}
In \cite{RUYZ15} it has been shown that the quantity $F_{FP}$ can be computed exactly in the infinite dimensional limit in terms of a mean square displacement matrix
\beq
\Delta_{ab}=\frac{d}{N}\sum_{i=1}^N|y_i^a-y_i^b|^2
\eeq
where $y_i^a=r_i^a$ if $a=1\ldots, m$ and $y_i^a=x_i^a$ if $a=m+1, \ldots, m+s$.
It has been shown that this is given by
\beq\label{eq:gauss_r}
-\b_s F_{FP}=s[\hat \a] = 1 - \log\r + d \log (m+s) + \frac{(m+s-1)d}{2} \log(2 \pi e D_g^2/d^2) + \frac{d}2 \log \det(\hat \a^{m+s,m+s}) -  \frac{d}2 \wh \f_g \,
 \FF\left( 2 \hat \a \right) \ ,
\eeq
where the matrix $\hat \alpha$ is a symmetric matrix whose elements are defined implicitly in terms of the matrix $\hat \Delta$ as
\beq
\Delta_{ab}=\alpha_{aa}+\a_{bb}-2\a_{ab}\:.
\eeq
We will call $s[\hat \a]$ the replicated entropy\footnote{The reader should not confuse the replicated entropy $s$ with the number of replicas $s$ that characterize the slave system. We will not choose to call them in a different way since it is quite straightforward to identify within a given context what the symbol $s$ refers to.}.
Moreover we will call
\beq
\frac{2}{d}s_{entr}=\log \det( \hat \a^{m+s,m+s})
\eeq
the ``entropic term'' of the replicated entropy. {Its expression} is given in terms of a matrix $\hat \a^{m+s,m+s}$ which is obtained from $\hat \a$ by removing the last column and row.
Moreover we define the ``interaction term'' as
\beq
\frac{2}{d}s_{int}\equiv-\wh \varphi_g \mathcal F(\hat \Delta)=-\wh \varphi_g \int_{-\infty}^\infty \frac{\de \z}{\sqrt{2\pi}} e^{-\z^2/2}\mathcal F_0\left(\Delta_{ab}+\frac{\z^2}{2}(\g_a-\g_b)^2\right)
\eeq
where
\beq\label{eq:FF0binomial}
\FF_0( \hat\D_{ab} ) \equiv \lim_{n\to 0}\sum_{n_1,\ldots, n_k; \sum_{a=1}^k n_a=n}\frac{n!}{n_1!\ldots n_k!}
e^{\sum_{a=1}^k \frac{n_a}{n} \h_a- \frac{1}{2} \sum_{a,b}^{1,k}\frac{n_an_b}{n^2}  \D_{ab}  } \ .
\eeq
and $\eta_a=\g_a=0$ for $a=1,\ldots m$ and $\eta_a=\eta$ and $\g_a=\g$ for $a=m+1,\ldots, m+s$.
In order to obtain the thermodynamical value of $F_{FP}$ we need to take the saddle point equations over the matrix $\hat \Delta$ \cite{KPZ12}.
This is not possible for a generic MSD matrix. The reason is that, besides the fact that the saddle point equations for the single entries $\Delta_{ab}$ are not very tractable, we have to take at the end an analytic continuation to non integer values of $s$.\\ 
The solution of the problem comes from symmetry (and symmetry breaking) arguments.
One can try to parametrize simply the matrix $\hat\D$. Its general form is given by
\beq
\hat \Delta=\begin{pmatrix}
\hat\Delta^g & \hat \Delta^r\\
(\hat\Delta^r)^T & \hat\Delta^s
\end{pmatrix}.
\eeq
$\hat \Delta^g$ is a $m\times m$ matrix that describes the MSDs of the master replicas while $\hat \Delta^s$ is a $s\times s$ matrix that encodes for the MSDs of the slave $s$ replicas. Finally $\hat \Delta^r$ is the matrix that gives the MSDs between master and slave replicas. 
In \cite{RUYZ15} we have considered a simple parametrization for them that we call a 1RSB parametrization. It corresponds to 
\beq
\begin{split}
\Delta^g_{ab}&=\Delta^g(1-\delta_{ab}) \ \ \ \ \ \ \forall\  a,b=1,\ldots, m\\
\Delta^s_{ab}&=\Delta(1-\delta_{ab}) \ \ \ \ \ \ \ \forall\  a,b=1,\ldots, s\\
\Delta^r_{ab}&=\Delta^r \ \ \ \ \ \ \ \ \ \ \ \ \ \ \ \ \ \forall a,b
\end{split}
\label{1RSBpara}
\eeq
Once this parametrization is chosen, we can obtain the thermodynamic potential $F_{FP}$ by taking the saddle point over $\Delta^r$, $\Delta^g$ and $\Delta$.
Then, the validity of the {calculation} must be checked by looking at the stability of the saddle point solution.
In \cite{RUYZ15} we have shown that for sufficiently strong compressions or strains (namely for high enough $\eta$ or $\g$) this parametrization is unstable and we need to go beyond it.
In particular we have shown that the replicon eigenvalue \cite{dAT78}, that is responsible for the replica symmetry breaking in the sector {of the slave replicas, becomes unstable} \cite{RUYZ15}. 
To correct this instability we need to consider a {more complicated} parametrization for the matrix $\hat \Delta$. This is called a full-replica-symmetry-breaking (fullRSB) ansatz \cite{MPV87}. 
In order to see how it is constructed we first notice that the MSD matrix for the master replicas cannot be affected by the slave replicas. 
This means that the saddle point equations for $\hat \Delta^g$ are completely independent on $\hat \Delta^r$ and $\hat \Delta^s$.
In this work we focus on the case wherein the master replicas are taken at equilibrium, $m=1$ as already discussed. In \cite{KPUZ13,RUYZ15} it has been shown that on this equilibrium line {the 1RSB ansatz is always stable}, which means that the 1RSB parametrization is still correct for the matrix $\hat \D^g$. Conversely, since the replicon instability is only in the slave $s$-sector we need to consider a fullRSB ansatz restricted to the sub-matrix $\hat \Delta^s$
\beq
\begin{split}
\hat \Delta^s&\to \{0, \Delta(x)\}\:.
\end{split}
\eeq
that is encoded in the continuous function $\Delta(x)$ with $x\in[0,1]$. Here we do not discuss in details how the fullRSB parametrization is {mathematically handled} since this can be exhaustively found in \cite{CKPUZ14JSTAT, MPV87}.\\
We will now derive the saddle point equations for $\Delta^r$ and the fullRSB profile $\Delta(x)$. To do this we need first to evaluate the replicated entropy on the fullRSB ansatz, {which we do below. We will consider} separately the entropic and interaction terms.

\subsection{Entropic term}

The entropic term for the replicated entropy has the following definition in terms of the overlap matrix $\alpha_{ab}\equiv \la x_a\cdot x_b\ra$ and $\hat \Delta$ \cite{KMZ15}:
\beq
\frac 2d s_{entr} \equiv \log\det\hat \alpha^{m+s,m+s}= \log\left[-\frac{2}{(m+s)^2}\left(\sum_{ab}(\hat\Delta)^{-1}_{ab}\right)\det(-\hat\Delta/2)\right].
\eeq
Let us now compute separately the two terms
\beq
\log \det \hat \Delta \ \ \ \ \ \ \ \ \ \ \ \ \ \ \ \sum_{a,b=1}^{m+s}\left[\hat\Delta^{-1}\right]_{ab}\:.
\label{twoterms}
\eeq
Let us start from the first one that can be rewritten as
\beq
\det \hat \Delta=\left(\det \hat\D^g\right)\det \left(\hat \D^s-\hat\Delta^r (\hat\Delta^{g})^{-1} (\hat \Delta^r)^T\right)\:.
\eeq
Using the fact that for a $m\times m$ replica symmetric matrix of the form $q_{ab}=\delta_{ab}+(1-\delta_{ab})q$ the determinant can be written as
\begin{equation}
\det  \hat q = (1-q)^{m-1} [ 1+(m-1)q ]
\end{equation}
we have that
\beq
\det \hat \D_q = \lim_{\epsilon\to 0}\left(\hat \Delta^g+\epsilon \mathbb 1_{m}\right)=\lim_{\epsilon\to 0}\epsilon^m \left(1-\frac{\Delta^g}{\epsilon}\right)^{m-1} \left[ 1+(m-1)\frac{\Delta^g}{\epsilon} \right]= (1-m)(-\Delta^g)^m
\label{detdeltag}
\eeq
where $\mathbb 1_m$ is the $m$ dimensional identity matrix.
Moreover we have
\beq
(\Delta^g)^{-1}_{ab} = -\frac{1}{\Delta^g}\left(\delta_{ab}+\frac{1}{1-m}\right)
\eeq
so that
\beq
\left[\hat\Delta^r (\hat\Delta^{g})^{-1} (\hat \Delta^r)^T\right]_{ab}=-\frac{(\Delta^r)^2}{\Delta^g}\frac{m}{1-m}
\eeq
This means that the matrix $\hat \Omega=\hat\Delta^s-\hat\Delta^r (\hat\Delta^{g})^{-1} (\hat \Delta^r)^T$ will be parametrized within the fullRSB ansatz by
\beq
\hat \Omega\to \left\{\Omega_d, \Omega(x)\right\}\ \ \ \ \ \ \ \ \ x\in[s,1]
\eeq
where
\beq
\begin{split}
\Omega_d&=\frac{(\Delta^r)^2}{\Delta^g}\frac{m}{1-m}\\
\Omega(x)&=\Delta(x)+\frac{(\Delta^r)^2}{\Delta^g}\frac{m}{1-m}\:.
\end{split}
\eeq
In this way we can use the result of \cite{MP91, CKPUZ14JSTAT} to obtain
\beq
\log\det \hat \Omega=  s\log(\Omega_d - \la \Omega \ra)  - s \int_s^1 \frac{\de y}{y^2} \log\left(\frac{
\Omega_d - \la \Omega \ra - [\Omega](y)}{\Omega_d-\la \Omega\ra}
\right)
\label{detomega}
\eeq
where
\beq
[\Omega](x)=x \Omega(x) - \int_0^x\de y\, \Omega(y) \ , 
\hskip2cm
 \langle \Omega \rangle = \int_0^1\de x\, \Omega(x) \ ,
\eeq
and we are assuming $\Omega(x)=0\ \ \forall x<s$.
By inserting the fullRSB parametrization for $\hat \Omega$ we get the computation of the first term of (\ref{twoterms}). Now we turn to the computation of the second term of (\ref{twoterms}).
We need to compute the inverse of the matrix $\hat\Delta$. We thus consider a general matrix of the following form
\beq
\hat G=\left(
\begin{array}{cc}
\hat q_g & \hat q_r^{(1)} \\
\hat q_r^{(2)} & \hat q\\
\end{array}
\right)
\eeq
where the fullRSB structure is only inside the sub-block $\hat q\to \left\{\tilde q, q(x)\right\}$ and the entries of the matrices $\hat q_r^{(1)}$ and $\hat q_r^{(2)}$ are all equal respectively to $q_r^{(1)}$ and $q_r^{(2)}$. The matrix $\hat q_g$ is parametrized by $\hat q_g\to\left\{q_d, g(x)\right\}$ and $g(x)=q_g$ if $x\in [m,1]$ and zero otherwise. Finally we will always assume that $q(x)=0$ for $x\in[0,s]$. 
We want to solve the inverse matrix problem, namely we want to find the matrix 
\beq
\hat G^{-1}=\left(
\begin{array}{cc}
\hat p_g & \hat p_r^{(1)} \\
\hat p_r^{(2)} & \hat p\\
\end{array}
\label{inverse_general}
\right)
\eeq
such that $\hat G\hat G^{-1}=\mathbb 1_{m+s}$. 
We search for a form of the inverse matrix $G^{-1}$ that is parametrized as follows.
The off diagonal blocks $\hat p_r^{(1)}$ and $\hat p_r^{(2)}$ are constant matrices whose entries are all equal respectively to $p_r^{(1)}$ and $p_r^{(2)}$.
Moreover we consider the parametrization $\hat p_g\to \{p_d, \g(x)\}$ where $\g(x)=p_g$ for $x\in[m,1]$ and $\g(x)=0$ otherwise.
Finally the slave replicas sector is parametrized by $\hat p\to \{\tilde p, p(x)\}$ with $p(x)=0$ for $x\in[0,s]$.
Using {these parametrizations}, the equations for the inverse of the matrix $G$ are
\begin{eqnarray}
&q_dp_d+(m-1)q_gp_g+sq_r^{(1)}p_r^{(2)}=1 \label{inverse_matrix_1}\\
&q_dp_g+q_g p_d+(m-2)q_g p_g+sq_r^{(1)} p_r^{(2)}=0\label{inverse_matrix_2}\\
&q_r^{(2)}p_d+(m-1)p_gq_r^{(2)}+p_r^{(2)}\left(\tilde q-\la q\ra\right)=0\label{inverse_matrix_3}\\
&q_d p_r^{(1)}+(m-1)q_g p_r^{(1)}+q_r^{(1)}\left(\tilde p-\la p\ra \right)=0 \label{inverse_matrix_4}\\
&mq_r^{(2)}p_r^{(1)}+\tilde q\tilde p-\int_s^1\de x q(x)p(x)=1\label{inverse_matrix_5}\\
&mq_r^{(2)}p_r^{(1)}-sp(x)q(x)+(\tilde q-\la q\ra)p(x)+(\tilde p-\la p\ra)q(x)-\int_s^x\de y(q(y)-q(x))(p(y)-p(x))=0\label{inverse_matrix_6}
\end{eqnarray}
These equations can be solved exactly. Let us focus on the last two equations and let us call $A(x)$ the right hand side of eq. (\ref{inverse_matrix_6}). This equation holds for all $x$ in the interval $[0,1]$. If we consider its derivative {with respect to $x$} we get
\beq
0=\dot A(x)= (\tilde p-\la p\ra)\dot q(x)+(\tilde q-\la q\ra)\dot p(x)-\dot p(x) [q](x)-\dot q(x)[p](x)
\eeq
Let us now consider the following quantity:
\beq
B(x)=\left(\tilde p-\la p \ra-[p](x)\right)\left(\tilde q-\la q \ra-[q](x)\right)
\eeq
It is simple to show that $\dot A(x)=-x\dot B(x)$ so that we obtain
\beq
\left(\tilde p-\la p \ra-[p](x)\right)\left(\tilde q-\la q \ra-[q](x)\right)=\aleph
\label{pqu1}
\eeq
where $\aleph$ is {independent of $x$}.
Computing (\ref{pqu1}) in $x=1$ and using eq. (\ref{inverse_matrix_5}) we get
\beq
\left(\tilde p-p(1)\right)\left(\tilde q-q(1)\right)=\aleph
\eeq
Moreover let us consider again eq. (\ref{inverse_matrix_6}) evaluated in $x=1$.
We get
\beq
1=\left(\tilde p-p(1)\right)\left(\tilde q-q(1)\right)
\eeq
so that we have $\aleph=1$.
Let us consider again the equation (\ref{pqu1}) evaluated in $x=s$. We get
\beq
\tilde p-\la p\ra-sp(s)=\frac{1}{\tilde q-\la q \ra-sq(s)}
\eeq
Using again eq. (\ref{inverse_matrix_6}) evaluated at $x=s$ we get
\beq
0=mp_r^{(1)}q_r^{(2)}-sp(s)q(s)+(\tilde p-\la p \ra)q(s)+(\tilde q-\la q \ra)p(s)
\eeq
By solving the last two equations with respect to $\tilde p-\la p \ra$ and $p(s)$ we get
\beq
\begin{split}
\tilde p -\la p \ra&=-\frac{s}{\tilde q-\la q \ra}\left[mp_r^{(1)}q_r^{(2)}\right]+\frac{1}{\tilde q-\la q \ra}\label{ptildeminavp}\\
p(s)&=-\frac{1}{\tilde q-\la q \ra}\left[mp_r^{(1)}q_r^{(2)}+\frac{q(s)}{\tilde q -\la q \ra-sq(s)}\right]
\end{split}
\eeq
from which we obtain
\beq
[p](x)=-\frac{[q](x)}{(\tilde q-\la q \ra)(\tilde q-\la q \ra-[q](x))}-\frac{smp_r^{(1)}q_r^{(2)}}{\tilde q-\la q \ra}\:.
\eeq
Taking the derivative with respect to $x$ we get
\beq
\dot p(x)=-\frac 1x \frac{\de}{\de x}\frac{[q](x)}{(\tilde q-\la q \ra)(\tilde q-\la q \ra-[q](x))}
\eeq
so that we have
\beq
\begin{split}
p(x)&=p(s)-\int_s^x\de y \frac 1y \frac{\de}{\de y}\frac{[q](y)}{(\tilde q-\la q \ra)(\tilde q-\la q \ra-[q](y))}\\
&=-\frac{1}{\tilde q -\la q \ra}\left[mp_r^{(1)}q_r^{(2)}+\frac 1x\frac{[q](x)}{\tilde q-\la q \ra-[q](x)}+\int_s^x \frac{\de y}{y^2}\frac{[q](y)}{\tilde q-\la q \ra  - [q](y)}   \right]
\end{split}
\eeq
and finally
\beq
\begin{split}
\la p\ra&=-mp_r^{(1)}q_r^{(2)}\frac{1-s}{\tilde q-\la q \ra}-\frac{1}{\tilde q - \la q \ra}\int_s^1\frac{\de y}{y^2}\frac{[q](y)}{\tilde q-\la q \ra-[q](y)}\\
\tilde p&=\frac{1}{\tilde q-\la q \ra}\left[1-mp_r^{(1)}q_r^{(2)}-\int_s^1\frac{\de y}{y^2}\frac{[q](y)}{\tilde q-\la q \ra-[q](y)}\right]\:.
\end{split}
\eeq
Let us now go back to the first four equations (\ref{inverse_matrix_1}-\ref{inverse_matrix_4}). We can use eq. (\ref{inverse_matrix_4}) together with \eqref{ptildeminavp} to solve for $p_r^{(1)}$. Finally the remaining three equations (\ref{inverse_matrix_1})-(\ref{inverse_matrix_3}) can be used to get $p_d$, $p_g$ and $p_r^{(2)}$:
\beq
\begin{split}
p_r^{(1)}&=-\frac{q_r^{(1)}}{q_d+(m-1)q_g}\left(\tilde q -\la q \ra-\frac{smq_r^{(1)}q_r^{(2)}}{q_d+(m-1)q_g}\right)^{-1}\\
p_r^{(2)}&=-\frac{q_r^{(2)}}{q_d+(m-1)q_g}\left(\tilde q -\la q \ra-\frac{smq_r^{(1)}q_r^{(2)}}{q_d+(m-1)q_g}\right)^{-1}\\
p_g&=-\frac{1}{q_d+(m-1)q_g}\left[\frac{q_g}{q_d-q_g}-\frac{sq_r^{(1)}q_r^{(2)}}{q_d+(m-1)q_g}\left(\tilde q -\la q \ra-\frac{smq_r^{(1)}q_r^{(2)}}{q_d+(m-1)q_g}\right)^{-1}\right]\\
p_d&=\frac{1}{q_d-q_g}-\frac{1}{q_d+(m-1)q_g}\left[\frac{q_g}{q_d-q_g}-\frac{sq_r^{(1)}q_r^{(2)}}{q_d+(m-1)q_g}\left(\tilde q -\la q \ra-\frac{smq_r^{(1)}q_r^{(2)}}{q_d+(m-1)q_g}\right)^{-1}\right]
\end{split}
\label{inverse_final_1}
\eeq
Note that the inverse of a symmetric matrix ($q_r^{(1)} = q_r^{(2)}$) is symmetric as well.
By inserting the expression of $p_r^{(1)}$ inside $\tilde p$ and $p(x)$ we end up with
\beq
\begin{split}
\tilde p&=\frac{1}{\tilde q-\la q \ra}\left[1+mq_r^{(2)}\frac{q_r^{(1)}}{q_d+(m-1)q_g}\left(\tilde q -\la q \ra-\frac{smq_r^{(1)}q_r^{(2)}}{q_d+(m-1)q_g}\right)^{-1}-\int_s^1\frac{\de y}{y^2}\frac{[q](y)}{\tilde q-\la q \ra-[q](y)}\right]\\
p(x)&=-\frac{1}{\tilde q -\la q \ra}\left[-mq_r^{(2)}\frac{q_r^{(1)}}{q_d+(m-1)q_g}\left(\tilde q -\la q \ra-\frac{smq_r^{(1)}q_r^{(2)}}{q_d+(m-1)q_g}\right)^{-1}\right.\\
&\left.+\frac 1x\frac{[q](x)}{\tilde q-\la q \ra-[q](x)}+\int_s^x \frac{\de y}{y^2}\frac{[q](y)}{\tilde q-\la q \ra  - [q](y)}   \right]
\end{split}
\label{inverse_final_2}
\eeq
and this completes the calculation of the inverse.
We can now collect all the results. Firstly we have that 
\beq
\log\det(\hat\Delta)= \log\det(\hat\Delta^g)+\log\det(\hat\Omega)\:.
\label{entropic1}
\eeq
Using \eqref{detdeltag}, the first term of the previous equation is easy and we get
\beq
\log\det\hat\Delta^g = \log(1-m) + m \log (-\Delta^g).
\eeq
To evaluate the second term of the right hand side of eq (\ref{entropic1}), we use the expression \eqref{detomega}. We have
\beq
\begin{split}
\Omega_d-\la\Omega\ra =\ &s\frac{(\Delta^r)^2}{\Delta^g}\frac{m}{1-m}-\la\Delta\ra,\\
\left[\Omega\right](y) =\ & s\frac{(\Delta^r)^2}{\Delta^g}\frac{m}{1-m} + \left[\Delta\right](y),
\end{split}
\eeq
so that 
\beq
\log\det(\hat \Omega) = s\log\left(s\frac{(\Delta^r)^2}{\Delta^g}\frac{m}{1-m} - \la\Delta\ra \right) - s\int_s^1 \frac{\de y}{y^2}\log\left(\frac{\la\Delta\ra+[\Delta](y)}{\la\Delta\ra - s\frac{(\Delta^r)^2}{\Delta^g}\frac{m}{1-m}}\right).
\eeq
We now need to perform the sum of the elements of $\hat\Delta^{-1}$. By parametrizing the entries of $\hat \Delta^{-1}$ with the same form of (\ref{inverse_general}) we get
\beq
\sum_{a,b=1}^{m+s}\left[\hat\Delta^{-1}\right]_{ab}  = msp_r^{(1)} + msp_r^{(2)} + mp_d + m(m-1)p_g + s(\tilde p -\la p \ra),
\label{secondtermexp}
\eeq
Using eqs. (\ref{inverse_final_1}) and (\ref{inverse_final_2}) with
\beq
\begin{split}
q_d =\ &0,\\
q_g =\ &\Delta^g,\\
q_r^{(1)} =\ &q_r^{(2)} = \Delta_r,\\
\tilde q = \ & 0,\\
q(x) =\ &\Delta(x),\\
\end{split}
\eeq
we get
\beq
msp_r^{(1)} + msp_r^{(2)}=2ms\left[\frac{\Delta^r}{(m-1)\Delta^g}\left(\la \Delta \ra + s\frac{(\Delta^r)^2}{\Delta^g}\frac{m}{m-1}\right)^{-1}\right],
\eeq
while
\beq
mp_d = -\frac{m}{\Delta^g} -\frac{m}{(m-1)\Delta^g}\left[-1+\frac{s(\Delta^r)^2}{(m-1)\Delta^g}\left(\la \Delta \ra + s\frac{(\Delta^r)^2}{\Delta^g}\frac{m}{m-1}\right)^{-1}\right],
\eeq
and
\beq
m(m-1)p_g=-\frac{m(m-1)}{(m-1)\Delta^g}\left[-1+\frac{s(\Delta^r)^2}{(m-1)\Delta^g}\left(\la \Delta \ra + s\frac{(\Delta^r)^2}{\Delta^g}\frac{m}{m-1}\right)^{-1}\right].
\eeq
The last term is then given by 
\beq
s(\tilde p -\la p \ra)=-\frac{s}{\la \Delta \ra + s\frac{(\Delta^r)^2}{\Delta^g}\frac{m}{m-1}}.
\eeq
Summing everything we obtain
\beq
\sum_{a,b=1}^{m+s}\left[\hat\Delta^{-1}\right]_{ab} = \frac{m}{(m-1)\Delta^g} -s\left(\frac{m\Delta^r}{(m-1)\Delta^g} -1\right)^{2}\left(\la \Delta \ra + s\frac{(\Delta^r)^2}{\Delta^g}\frac{m}{m-1}\right)^{-1}.
\eeq
The final expression of the entropic term for the replicated entropy is thus given by
\beq
\begin{split}
\frac 2d s_{entr}=\ &(1-m-s)\log2-2\log (m+s) + \log\left[\frac{m}{(1-m)\Delta^g} +s\left(\frac{m\Delta^r}{(m-1)\Delta^g} -1\right)^{2}\left(\la \Delta \ra - s\frac{(\Delta^r)^2}{\Delta^g}\frac{m}{1-m}\right)^{-1}\right]\\
&+\log(1-m) + m\log(\Delta^g) + s\log\left(\la\Delta\ra - s\frac{(\Delta^r)^2}{\Delta^g}\frac{m}{1-m} \right) - s\int_s^1 \frac{\de y}{y^2}\log\left(\frac{\la\Delta\ra+[\Delta](y)}{\la\Delta\ra - s\frac{(\Delta^r)^2}{\Delta^g}\frac{m}{1-m}}\right)=\\
=\ &(1-m-s)\log2-2\log (m+s) + (m-1)\log\Delta^g + \log[m\la\Delta\ra + 2ms\Delta^r + (1-m)s\Delta^g] \\
&- s\int_s^1 \frac{\de y}{y^2}\log\left[\la\Delta\ra+[\Delta](y)\right]
\end{split}
\label{entropic_final}
\eeq
This completes the calculation of the entropic term.
To check the validity of this expression we can look at two extreme cases.
First, we can consider the limit $s\to 0$.
In this case the matrix $\hat \Delta$ contains only the master replica sector and the replicated entropy becomes the same as the one considered in \cite{KPZ12, KPUZ13, CKPUZ14JSTAT}. Its entropic term is thus given by the expression
\beq
\log\det\alpha^{m,m} =\log\left[ \frac{2^{1-m}}{m^2}\frac{m}{(1-m)\Delta^g}(1-m)(\Delta^g)^{m}\right] = (m-1)\log\left(\frac{\Delta^g}{2}\right)-\log m
\eeq
that coincides with the results of \cite{KPZ12, KPUZ13, CKPUZ14JSTAT}.
The second way to check expression (\ref{entropic_final}) is to consider a replica symmetric profile for the slave sector.
This corresponds to impose a RS ansats for the $s$ replicas ($\Delta(x) = const = \Delta$) as done in \cite{RUYZ15}.
Doing this we get
\beq
\log\det\alpha^{m,m} = (1-m-s)\log2-2\log (m+s) + (m-1)\log\Delta^g + (s-1)\log\Delta + \log[ms\Delta^f + s\Delta^g + m\Delta],
\eeq
that coincides with the result of \cite{RUYZ15}.

\subsection{Interaction term}
The general term we need to compute is the following \cite{RUYZ15}
\beq\label{eq:FF0binomial}
\FF_0( \hat\D )  = \lim_{n\to 0}\sum_{n_1,\ldots, n_k; \sum_{a=1}^k n_a=n}\frac{n!}{n_1!\ldots n_k!}
e^{\sum_{a=1}^k \frac{n_a}{n} \h_a- \frac{1}{2} \sum_{a,b}^{1,k}\frac{n_an_b}{n^2}  \D_{ab}  } \ .
\eeq
By introducing Gaussian integrals, we can rewrite this term as \cite{CKPUZ14JSTAT}
\beq\label{int_gen}
\begin{split}
\mathcal F_0(\hat\Delta) 
= & \int_{-\infty}^\infty \de h\, e^{h} 
 \frac{\de}{\de h} \left\{
\exp\left[-\frac{1}{2}\sum_{a,b=1}^k \Delta_{ab}\frac{\partial^2}{\partial h_a\partial h_b}\right] \prod_{a=1}^{m+s}  \th(h_a)  
 \right\}_{\{h_a=h-\eta_a\}} \ . 
\end{split} 
\eeq
where $k=m+s$ and $\theta(x)$ is the step Heaviside function. Now we assume that the $s$-sector of the displacement matrix has a fullRSB structure.
Thus we have
\beq
\begin{split}
\mathcal F_0(\hat\Delta) 
=\ &\int_{-\infty}^\infty \de h\, \eee^{h} 
 \frac{\de}{\de h} \left\{
\exp\left[-\frac{1}{2}\sum_{a,b=1}^m \Delta_{ab}\frac{\partial^2}{\partial h_a\partial h_b}-\frac 12\sum_{a=1}^m\sum_{b=m+1}^{m+s}\Delta_{ab}\frac{\partial^2}{\partial h_a\partial h_b}-\frac 12\sum_{a=m+1}^{m+s}\sum_{b=1}^{m}\Delta_{ab}\frac{\partial^2}{\partial h_a\partial h_b}\right.\right.\\
&\left.\left.-\frac 12\sum_{a,b=m+1}^{m+s}\Delta_{ab}\frac{\partial^2}{\partial h_a\partial h_b}\right] \prod_{a=1}^k  \th(h_a)  
 \right\}_{\{h_a=h-\eta_a\}}=\\
=\ &\int_{-\infty}^\infty \de h\, \eee^{h} 
 \frac{\de}{\de h} \left\{
\exp\left[-\frac{1}{2}\Delta_g\left(\sum_{a=1}^m \frac{\partial}{\partial h_a}\right)^2+\frac{\Delta_g}{2}\sum_{a=1}^m\frac{\partial^2}{\partial h_a^2}-\Delta_r\left(\sum_{a=1}^m\frac{\partial }{\partial h_a}\right)\left(\sum_{b=m+1}^{m+s}\frac{\partial}{\partial h_b}\right)\right.\right.\\
&\left.\left.-\frac 12\sum_{a,b=m+1}^{m+s}\Delta_{ab}\frac{\partial^2}{\partial h_a\partial h_b}\right] \prod_{a=1}^k  \th(h_a)  
 \right\}_{\{h_a=h-\eta_a\}}\\
=&\ \int_{-\infty}^\infty \de h\, \eee^{h} 
 \frac{\de}{\de h} \left\{
\exp\left[-\frac{1}{2}\Delta_g\left(\sum_{a=1}^m \frac{\partial}{\partial h_a}\right)^2-\Delta_r\left(\sum_{a=1}^m\frac{\partial }{\partial h_a}\right)\left(\sum_{b=m+1}^{m+s}\frac{\partial}{\partial h_b}\right)\right.\right.\\
&\left.\left.-\frac 12\sum_{a,b=m+1}^{m+s}\Delta_{ab}\frac{\partial^2}{\partial h_a\partial h_b}\right]\left(\prod_{a=1}^{m}\Theta\left(\frac{h_a}{\sqrt{2\Delta_g}}\right)\right) \left(\prod_{b=m+1}^{m+s}  \th(h_b)\right)  
 \right\}_{\{h_a=h-\eta_a\}}\\
=&\ \int_{-\infty}^\infty \de h\, \eee^{h} 
 \frac{\de}{\de h} \left\{
\exp\left[-\frac{1}{2}\Delta_g\frac{\partial^2}{\partial h'^2}-\Delta_r\frac{\partial }{\partial h'}\frac{\partial}{\partial h''}-\frac 12\D_1\frac{\partial^2}{\partial h''^2}\right]\left(\Theta\left(\frac{h'}{\sqrt{2\Delta_g}}\right)\right)^m g^{s/s_1}(s_1,h''-\eta)  
 \right\}_{h'=h''=h}
\end{split} 
\label{interaction_zero}
\eeq
where the function $g(x,h)$ is {defined in terms of} $f(x,h)$ as
\beq
f(x,h) \equiv \frac 1x \log g(x,h)
\eeq
and satisfies the partial differential equation \cite{Pa79, Pa80, MPV87,Du81}
\beq
\frac{\partial f}{\partial x}=\frac 12\frac{\de \Delta(x)}{\de x}\left[\frac{\partial^2 f}{\partial h^2}+x\left(\frac{\partial f}{\partial h}\right)^2\right]
\label{Parisi_eq}
\eeq
with the initial condition
\beq
f(1,h)=\log \Theta \left(\frac{h}{\sqrt{2\Delta(1)}}\right)
\label{Parisi_initial}
\eeq
and as usual \cite{KPUZ13,CKPUZ14JSTAT, RUYZ15} we have defined
\beq
\Theta(x)=\frac{1}{2}\left(1+\erf(x)\right)\:.
\eeq
Note that we have defined $\Delta(s)=\Delta_1$ while $s_1$ is the smallest breaking point of the fullRSB profile $\Delta(x)$ (see \cite{CKPUZ14JSTAT} for details).
At this point we can manipulate the last expression to do the final integrals by parts and to give an integral representation for the exponential of differential operators.
We consider
\beq
\hat O=-\frac{1}{2}\Delta_g\frac{\partial^2}{\partial h'^2}-\Delta_r\frac{\partial }{\partial h'}\frac{\partial}{\partial h''}-\frac 12\D_1\frac{\partial^2}{\partial h''^2}
\eeq
and we introduce the differential operator
\beq
\hat H=\frac{\partial}{\partial h'}+\frac{\partial}{\partial h''}\:.
\eeq
We have
\beq
\hat O=\frac 12 \Delta_f \left(\frac{\partial}{\partial h''}\right)^2-\frac 12(\D_f+\D_1)\hat H\frac{\partial }{\partial h''}-\frac{\D_g}{2}\hat H\frac{\partial}{\partial h'}
\eeq
where we have defined $\D_f=2\D_r-\D_1-\D_g$ (see \cite{RUYZ15}).
By inserting this expression into the interaction term we get
\beq
\begin{split}
\mathcal F_0(\hat\Delta) =&\ \int_{-\infty}^\infty \de h\, \eee^{h} 
 \frac{\de}{\de h} \left\{
\exp\left[\frac 12 \Delta_f \left(\frac{\partial}{\partial h''}\right)^2-\frac 12(\D_f+\D_1)\hat H\frac{\partial }{\partial h''}-\frac{\D_g}{2}\hat H\frac{\partial}{\partial h'}\right]\right.\\
&\left.\times \left(\Theta\left(\frac{h'}{\sqrt{2\Delta_g}}\right)\right)^m g^{s/s_1}(s_1,h''-\eta)  
 \right\}_{h'=h''=h}\:.
 \end{split}
\eeq
Let us consider now a simple term of the form
\beq
\int_{-\infty}^\infty \de h\, \eee^{h} 
 \frac{\de}{\de h} \left\{\exp\left[A \hat H \frac{\partial }{\partial h''}\right]f(h',h'')\right\}_{h'=h''=h}=\left.\int_{-\infty}^\infty \de h\, \eee^{h}\frac{\de}{\de h} \sum_{k=0}^\infty\frac{1}{k!}A^k\hat H^k\frac{\partial^k}{\partial h''^k}f(h',h'')\right|_{h'=h''=h} \:.
\eeq
By integrating by parts all the terms of the series expansion we get
\beq
\begin{split}
&\int_{-\infty}^\infty \de h\, \eee^{h} 
 \frac{\de}{\de h} \left\{\exp\left[A \hat H \frac{\partial }{\partial h''}\right]f(h',h'')\right\}_{h'=h''=h}=\left.\int_{-\infty}^\infty \de h\, \eee^{h}\frac{\de}{\de h} \sum_{k=0}^\infty\frac{1}{k!}(-A)^k\frac{\partial^k}{\partial h''^k}f(h',h'')\right|_{h'=h''=h} \\
 &=\int_{-\infty}^\infty \de h\, \eee^{h} 
 \frac{\de}{\de h} \left\{\exp\left[-A \frac{\partial }{\partial h''}\right]f(h',h'')\right\}_{h'=h''=h}=\int_{-\infty}^\infty \de h\, \eee^{h} 
 \frac{\de}{\de h} \left\{f(h',h-A)\right\}_{h'=h''=h}.
\end{split}
\eeq
Using this result we finally get for the entropic term
\beq
\begin{split}
\mathcal F_0(\hat\Delta) &=\int_{-\infty}^\infty \de h\, \eee^{h} 
 \frac{\de}{\de h} \left\{
\exp\left[-\frac 12(\D_f+\D_1)\hat H\frac{\partial }{\partial h''}-\frac{\D_g}{2}\hat H\frac{\partial}{\partial h'}\right]\left(\Theta\left(\frac{h'}{\sqrt{2\Delta_g}}\right)\right)^m \gamma_{\D_f}\star g^{s/s_1}(s_1,h''-\eta) 
 \right\}_{h'=h''=h}=\\
&=\int_{-\infty}^\infty \de h\, \eee^{h} 
 \frac{\de}{\de h} \left\{
\left(\Theta\left(\frac{h+\D_g/2}{\sqrt{2\Delta_g}}\right)\right)^m \gamma_{\D_f}\star g^{s/s_1}(s_1,h-\eta+(\D_f+\D_1)/2) \right\}=\\
&=\int_{-\infty}^\infty \de h\, \eee^{h}  \left\{1-
\left(\Theta\left(\frac{h+\D_g/2}{\sqrt{2\Delta_g}}\right)\right)^m \gamma_{\D_f}\star g^{s/s_1}(s_1,h-\eta+(\D_f+\D_1)/2) \right\}\:.
 \end{split}
\eeq
where we have used that \cite{CKPUZ14JSTAT}
\beq
\eee^{\frac{a}{2}\frac{\partial^2}{\partial h^2}}f(h)=\g_{a}\star f(h)\ \ \ \ \ \ \ \ \ \ \ \ \g_{a}\star f(h)=\int_{-\infty}^\infty\frac{\de z}{\sqrt{2\pi a}}\eee^{-\frac{z^2}{2a}}f(h-z)\:.
\label{differential_gaussian}
\eeq

\subsection{Final result for the entropy of the $m+s$ replicas}
Now we can collect all the results of the previous section to write the final expression of the replicated entropy.
We have\footnote{We have rewritten the integral part of the entropic term in a way that ensures convergence of the integral in the limit $s\to 0$.}
\beq
\begin{split}
s[\hat\D] = &1 - \log\r + \frac{(m+s-1)d}{2} \log(\pi e D^2/d^2) + \frac{d(m-1)}{2}\log\Delta^g + \frac{d}{2}\log[m(\la\D\ra+s\D_1) + ms\D^f \\
&+ s\D^g] +\frac d2(s-1)\log\la\D\ra- s\frac{d}{2}\int_s^1\frac{\de y}{y^2}\log\left(\frac{\la\D\ra + [\D](y)}{\la\D\ra}\right)\\
&-\frac{d\wh\varphi_g}{2}\int_{-\infty}^\infty \frac{\de \zeta}{\sqrt{2\pi}}\eee^{-\zeta^2/2}\int_{-\infty}^\infty \de h\, \eee^{h}  \left\{1-\left(\Theta\left(\frac{h+\D_g/2}{\sqrt{2\Delta_g}}\right)\right)^m \gamma_{\D_\g(\z)}\star g^{s/s_1}(s_1,h-\eta+(\D_\g(\zeta)+\D_1)/2) \right\}
\end{split}
\label{final_replicated_entropy}
\eeq
where we have defined $\Delta_\g(\zeta)=\Delta_f+\zeta^2\g^2$.
Again, in order to check the expression above it is quite useful to look for known limiting cases. The first one is again the limit $s\to 0$ that gives the so called Monasson \cite{Mo95} replicated entropy 
\beq
\begin{split}
\lim_{s\to 0} s[\hat \a] = s_{\rm 1RSB}(\D) =\ &1 -\log\r + \frac{d}2 (m-1) + \frac{d}2 \log m + \frac{d}2 (m-1) \log( \pi \D/d^2 ) \\
& -\frac{d}2 \wh\f_g \int_{-\infty}^\infty \de y \, \eee^y \, \left[1 - \Th\left(  \frac{y + \D/2}{\sqrt{2\D}} \right)^m \right] \ .
\end{split}
\eeq
that has the same form as the one obtained in \cite{CKPUZ14JSTAT}.

Moreover we can again impose an RS ansatz in the slave replicas sector $\Delta(x)=\Delta$ for all $x\in[s,1]$. In this case the function $g$ is given by
\beq
g(s_1,h) = \Theta\left(\frac{h}{\sqrt{2\Delta}}\right)^{s_1}\:.
\eeq
It useful to change integration variable $-x+\Delta^f/2 = x'$ in the Gaussian convolution that appears in the interaction term to get (for simplicity we focus only on the case $\g=0$)
\beq
\begin{split}
s_{\textrm{1RSB}}[\a] = &1 - \log\r + \frac{(m+s-1)d}{2} \log(\pi e D^2/d^2) + \frac{d(m-1)}{2}\log\Delta^g + \frac{d}{2}\log[m\D + ms\D^f + s\D^g] + \frac{d(s-1)}{2}\log\Delta\\&-\frac{d\wh\varphi_g}{2}\int_{-\infty}^\infty \de h \, \eee^{h} \, \left\{  1 
- \Th\left(  \frac{h + \D^g/2}{\sqrt{2\D^g}} \right)^m \int_{-\infty}^\infty \de x' \,  \Th\left(  \frac{ x' + h - \eta +\D/2}{\sqrt{2 \D}} \right)^s 
\frac{ \eee^{- \frac1{2 \D^f} \left( x' - \D^f/2  \right)^2  } }{\sqrt{2\pi \D^f}} \right\},
\end{split}
\eeq
that is what we have obtained in \cite{RUYZ15}.

\subsection{Final result for the entropy of the planted state}
If we want to compute the free energy of the followed glassy states we have to plug the expression for the replicated entropy inside eq. (\ref{limit_sto0}). The simplest way to perform the limit $s\to 0$ is to develop the replicated entropy in powers of $s$ and take the linear order.
First of all we have
$$
g(s_1,h)^{s/s_1}\simeq 1+ \frac{s}{s_1}\log g(s_1,h) + O(s^2).
$$
Taking the $s\to 0$ limit we get the final result
\beq
\begin{split}
-\beta_s F_g= s_g[\a] =\ &\frac{d}{2} + \frac{d}{2}\log\left(\frac{\pi\la\D\ra}{d^2}\right) - \frac{d}{2}\int_0^1\frac{\de y}{y^2}\log\left(\frac{\la\D\ra + [\D](y)}{\la\D\ra}\right) + \frac{d}{2}\frac{m\D^f+\D^g}{m\la\D\ra}\\ 
&+\frac{d\wh\varphi_g}{2}\int_{-\infty}^\infty\DD \z \int_{-\infty}^\infty \de h \, \eee^{h}\Th\left(  \frac{h + \D^g/2}{\sqrt{2\D^g}} \right)^m \int_{-\infty}^\infty \de x' \,  f (0,x' + h - \eta +\D(0)/2) 
\frac{ \eee^{- \frac1{2 \D_\g(\z)} \left( x' - \D_\g(\z)/2  \right)^2  } }{\sqrt{2\pi \D_\g(\z)}}.
\end{split}
\label{eq:sfull}
\eeq
where we have defined $\DD \z=\de \z\eee^{-\z^2/2}/\sqrt{2\pi}$. 
This expression gives the entropy for a generic matrix $\hat \Delta$. In order to compute the Franz-Parisi entropy of the glass state, we need to
fix $\hat \Delta$ through the saddle point equations \cite{RUYZ15}.

\subsection{Simplifications for $m=1$\label{sec:m=1}}
Before {proceeding} with the variational equations for $\hat \Delta$ we want to show that in the case in which the master replicas are taken at equilibrium, namely when $m=1$, the form of the state-followed entropy can be much simplified. Indeed in this case $\Delta^g$ disappears from the equations.\\
It is quite easy to see this in the case of the entropic term of eq. (\ref{final_replicated_entropy}) by remembering that $\Delta^f \equiv 2\Delta^r-\Delta(s)-\Delta^g$.
It remains to verify that $\Delta^g$ disappears also from the interaction term. For $m=1$ its general form is given by 
$$
\mathcal{F}(\wh\D) =\int_{-\infty}^\infty\DD \z \int_{-\infty}^\infty \de y \, \eee^{y} \, \left\{  1 
- \Th\left(  \frac{y + \D^g/2}{\sqrt{2\D^g}} \right) \int_{-\infty}^\infty \de x \,  \s(x+y) 
\frac{ \eee^{- \frac1{2 \D_\g(\z)} \left( x - \D_\g(\z)/2  \right)^2  } }{\sqrt{2\pi \D_\g(\z)}} \right\},
$$
where in our case the function $\s(x)$ is given by
\beq
\s(x)=g^{s/s_1}\left(s_1,x-\eta+\frac{\Delta_1}{2}\right)
\label{sigmafullRSB}
\eeq
This general form is valid for every replica-symmetry-breaking ansatz (the only difference is in the specific form of $\s(x)$). We then express the $\Th$ function 
with its integral representation
\beq
\Theta\left(\frac{h+\Delta^g/2}{\sqrt{2\Delta^g}}\right)=\int_{-\infty}^\infty \frac{\de \l}{\sqrt{2\pi \Delta^g}}\eee^{-\l^2/(2\Delta^g)} \theta\left(h+\frac{\Delta^g}{2}-\l\right)
\label{Theta_rep}
\eeq
to get
\beq
\mathcal{F}(\wh\D) =\int_{-\infty}^\infty\DD\z \int_{-\infty}^\infty \de x\de y\de \l\ \eee^{y}\frac{\eee^{-\frac{(\l+\D^g/2)^2}{2\D^g}}}{\sqrt{2\pi\D^g}}\frac{\eee^{-\frac{(x-\D_\g(\z)/2)^2}{2\D_\g(\z)}}}{\sqrt{2\pi\D_\g(\z)}}[1-\theta(y-\l)\s(x+y)]\:.
\eeq 
We now change integration variables in the following way:
\beq
\begin{cases}
u = y+x\\
v = \l+x\\
w = x\:.
\end{cases}
\eeq
Note that the Jacobian of this change of coordinates is one so that we get
\beq
\mathcal{F}(\wh\D) = \int_{-\infty}^\infty\DD \z\int_{-\infty}^\infty \de u \de v  \de w\ \eee^{u-w}\frac{\eee^{-\frac{(v-w+\D^g/2)^2}{2\D^g}}}{\sqrt{2\pi\D^g}}\frac{\eee^{-\frac{(w-\D_\g(\z)/2)^2}{2\D_\g(\z)}}}{\sqrt{2\pi\D_\g(\z)}}[1-\theta(u-v)\s(u)].
\eeq
The integral on $w$ can be easily done analytically, since it is just a convolution of two Gaussians. We get
\beq
\mathcal{F}(\wh\D) = \int_{-\infty}^\infty\DD \z\int_{-\infty}^\infty \de u \de v\ \eee^{u}\frac{\eee^{-\frac{(v+\D^g/2+\D_\g(\z)/2)^2}{2(\D^g+\D_\g(\z))}}}{\sqrt{2\pi(\D^g+\D_\g(\z))}}[1-\theta(u-v)\s(u)]\:.
\eeq
Remembering that $\D_g+\D_f = 2\D_r-\D_1$, we get that $\D_g$ disappears from the expression. 
Using again (\ref{Theta_rep}) we get finally
\beq
\mathcal{F}(\wh\D) = \int_{-\infty}^\infty\DD \z\int_{-\infty}^\infty \de u\ \eee^{u}\left\{1-\Th\left(\frac{u+(2\D_r+\g^2\z^2-\D_1)/2}{\sqrt{2(2\D_r+\g^2\z^2-\D_1)}}\right)\s(u)\right\}.
\eeq
This expression is much simpler than the corresponding one with $m \neq 1$ and we will take advantage from it in the numerical solution of the saddle point equations for the MSD matrix.
Finally, {we recall that} with a 1RSB ansatz the function $\s$ would be given by
\beq
\s_{\textrm{1RSB}}(u) = \Th\left(\frac{u-\eta+\D_1/2}{\sqrt{2\D_1}}\right)^s,
\eeq
and for the more general fullRSB case it is given by (\ref{sigmafullRSB}).

\section{Variational equations}
Having computed the expression for the replicated entropy within a fullRSB ansatz for the slave replicas sector, we are equipped now to write the variational equations for the saddle point value of the matrix $\hat\Delta$.
In \cite{CKPUZ14JSTAT} this has been done in two different ways: firstly it has been shown how the fullRSB variational equations can be obtained from the finite $k$-RSB ones in the limit $k\to \infty$ and then it has been shown how to recover the corresponding results using Lagrange multipliers \cite{SD84}. Here we follow the same strategy.

\subsection{Finite $k$RSB equations}
We consider again the replicated entropy given in equation (\ref{final_replicated_entropy}). In order to write down the $k$RSB saddle point equations we need to rewrite the entropy in a generic $k$RSB ansatz. We firstly introduce a new quantity \cite{CKPUZ14JSTAT}
\beq
G_i=s_i\Delta_i+\sum_{j=i+1}^k(s_j-s_{j-1})\Delta_j\ \ \ \ \       \Longleftrightarrow\ \ \ \ \        \Delta_i=\frac{G_i}{s_i}+\sum_{j=i+1}^k\left(\frac{1}{s_j}-\frac{1}{s_{j-1}}\right)G_j
\eeq
that in the continuum limit becomes
\beq
G(x)=x\Delta(x)+\int_{x}^1\de z\Delta(z)\ \ \ \ \      \Longleftrightarrow\ \ \ \ \       \Delta(x)=\frac{G(x)}{x}-\int_x^1\frac{\de z}{z^2}G(z)\:.
\eeq
The replicated entropy evaluated on a $k$RSB ansatz can then be rewritten as
\beq
\begin{split}
\frac{2}{d}\SS_{k\textrm{RSB}}=\ &\textrm{const.}+(m-1)\log\Delta^g+\log\left[mG_1+ms \Delta^f+s\Delta^g\right] +\sum_{j=1}^k\left(\frac{s}{s_j}-\frac{s}{s_{j-1}}\right)\log G_j\\
&-\wh\f_g\int_{-\infty}^\infty \DD \z \int_{-\infty}^\infty \de h\, \eee^{h} 
 \frac{\de}{\de h} \left\{
\left(\Theta\left(\frac{h+\D_g/2}{\sqrt{2\Delta_g}}\right)\right)^m \gamma_{\D_\g(\z)}\star g^{s/s_1}(s_1,h-\eta+(\D_\g(\z)+\D_1)/2) \right\}\:.
\end{split}
\eeq
The saddle point equations are obtained by taking the derivatives of the replicated entropy with respect to $\Delta_r$ and $\Delta_i$\footnote{Note that the saddle point equation for $\Delta_g$ is already fixed \cite{RUYZ15}.}.
However, since the relation between $\Delta^r$ and $\Delta_i$ and $\Delta^f$ and $G_i$ is invertible, we choose to take the derivative with respect to $\Delta^f$ and $G_i$.
We start by taking the derivative with respect to $\Delta^f$. Let us consider the derivative of the entropic term of the replicated entropy. We get
\beq
\frac 2d \frac{\partial s_{entr}}{\partial \Delta^f}=\frac{ms}{mG_1+ms\Delta^f+s\Delta^g }.
\eeq
By taking the derivative of the interaction part we get
\beq
\begin{split}
\frac 2d \frac{\partial s_{int}}{\partial \Delta^f}=&\ \frac{\wh \f_g}{2} \int_{-\infty}^\infty\DD \z \int_{-\infty}^\infty \de h \eee^h \Theta^m\left[\frac{h+\Delta^g/2}{\sqrt{2\D^g}}\right]\g_{\Delta_\g(\z)}\star \left(\frac{\partial^2}{\partial h^2}+\frac{\partial}{\partial h}\right)g^{s/s_1}\left(s_1, h-\eta+\frac{\Delta_\g(\z)+\Delta_1}{2}\right)\\
=&\ \frac{\wh \f_g}{2} \int_{-\infty}^\infty\DD \z \int_{-\infty}^\infty \de h\left[\g_{\Delta_\g(\z)}\star\left(\eee^h \Theta^m\left[\frac{h+\Delta^g/2}{\sqrt{2\D^g}}\right]\right)\right]\eee^{sf(s_1,h-\eta+(\Delta_\g(\z)+\Delta_1)/2)}\\
&\times\left(sf'(s_1,h-\eta+(\Delta_\g(\z)+\Delta_1)/2)+sf''(s_1,h-\eta+(\Delta_\g(\z)+\Delta_1)/2)+\left((s f'(s_1,h-\eta+(\Delta_\g(\z)+\Delta_1)/2)\right)^2\right)
\end{split}
\label{interazione_Delta_f}
\eeq
that gives
\beq
\begin{split}
\frac{ms}{mG_1+ms\Delta^f+s\Delta^g }=&-\frac{\wh\f_g}{2}\int_{-\infty}^\infty\DD \z \int_{-\infty}^\infty \de h\left[\g_{\Delta_\g(\z)}\star\left(\eee^{h+\eta-(\Delta_\g(\z)+\Delta_1)/2} \Theta^m\left[\frac{h+\eta -(\Delta_\g(\z)+\Delta_1-\Delta^g)/2}{\sqrt{2\D^g}}\right]\right)\right]\\
&\times \eee^{sf(s_1,h)}\left(\left(sf'(s_1,h)\right)^2+sf'(s_1,h)+sf''(s_1,h)\right)\:.
\end{split}
\label{equazione_Delta_f_finite_s_prima}
\eeq
We can then define
\beq
P(s_1,h)\equiv\eee^{h+\eta-\Delta_1/2}\int_{-\infty}^\infty \DD \z \int_{-\infty}^\infty\frac{\de x}{\sqrt{2\pi \Delta_\g(\z)}}\eee^{-\frac{1}{2\Delta_\g(\z)}\left(x+\frac{\Delta_\g(\z)}{2}\right)^2} \Theta \left[\frac{h+\eta -z+(\Delta^g-\Delta_1)/2}{\sqrt{2\Delta^g}}\right]^m\eee^{sf(s_1,h)}
\label{defP1}
\eeq
so that Eq. (\ref{equazione_Delta_f_finite_s_prima}) becomes
\beq
\frac{ms}{mG_1+ms\Delta^f+s\Delta^g }=-\frac{\wh\f_g}{2}\int_{-\infty}^\infty\de h P(s_1,h)\left(\left(sf'(s_1,h)\right)^2+sf'(s_1,h)+sf''(s_1,h)\right)\:.
\label{equazione_Delta_f_finite_s}
\eeq
Taking the leading order for $s\to 0$ we get
\beq
\frac{1}{G_1}=-\frac{\wh \f_g}{2}\int_{-\infty}^\infty \de h P(s_1,h)\left(f''(s_1,h)+f'(s_1,h)\right)
\label{equazione_Delta_f}
\eeq
where here $P(s_1,h)$ is the $s\to 0$ limit of (\ref{defP1})
\beq
P(s_1,h)=\eee^{h+\eta-\Delta_1/2}\int_{-\infty}^\infty \DD \z \int_{-\infty}^\infty\frac{\de x}{\sqrt{2\pi \Delta_\g(\z)}}\eee^{-\frac{1}{2\Delta_\g(\z)}\left(x+\frac{\Delta_\g(\z)}{2}\right)^2} \Theta^{m}\left[\frac{h+\eta -x+(\Delta^g-\Delta_1)/2}{\sqrt{2\Delta^g}}\right]
\eeq
In the continuum limit, the equation above becomes
\beq
\frac{1}{G(0)}=-\frac{\wh \f_g}{2}\int_{-\infty}^\infty\de h\, P(0,h)\left(f''(0,h)+f'(0,h)\right)
\label{equazione_G0}
\eeq
with {$G_1=G(0)$}.
At this point, we can consider the derivatives with respect to $G_i$. 
Let us first consider the entropic term. We get
\beq
\frac 2d \frac{\partial s_{entr}}{\partial G_i}=\delta_{i,1}\frac{m}{m G_1+ms\Delta^f+s\Delta^g}+\left(\frac{s}{s_i}-\frac{s}{s_{i-1}}\right)\frac{1}{G_i}
\eeq
The interaction term instead gives
\beq
\begin{split}
\frac 2d \frac{\partial s_{int}}{\partial G_i}=\ &\wh \f_g \frac{\partial }{\partial \Delta^f}\int_{-\infty}^\infty \DD \z\int_{-\infty}^\infty \de h\, \eee^{h} 
 \frac{\de}{\de h} \left\{
\left(\Theta\left(\frac{h+\D_g/2}{\sqrt{2\Delta_g}}\right)\right)^m \gamma_{\D_\g(\z)}\star g^{s/s_1}(s_1,h-\eta+(\D_\g(\z)+\D_1)/2) \right\}\\
&\times\left(\frac{1}{s_1}\delta_{i,1}+(1-\delta_{i,1})\left(\frac{1}{s_i}-\frac{1}{s_{i-1}}\right)\right)\\
&+s\frac{\wh \f_g}{2}\int_{-\infty}^\infty\DD\z \int_{-\infty}^\infty\de h \left[\g_{\Delta_\g(\z)}\star \left(\eee^{h+\eta-(\D_\g(\z)+\D_1)/2}\Theta^m\left[\frac{h+\eta-(\D_\g(\z)+\D_1-\D^g)/2}{\sqrt{2\D^g}}\right]\right)\right]\\
&\times \left\{\frac{s_i-s_{i-1}}{s_i}\G_i \ast \left(f'(s_i,h)\right)^2+\left(\frac{1}{s_i}-\frac{1}{s_{i-1}}\right)\sum_{j=1}^{i-1}(s_j-s_{j-1})\G_j\ast\left(f'(s_j,h)\right)^2\right\}
\end{split}
\label{derivative_int}
\eeq
where as in \cite{CKPUZ14JSTAT} we have defined the operators $\G_l$ that satisfy the following recursion relations
\beq
\begin{split}
\G_1\ast t(h)&=\eee^{sf(s_1,h)}t(h)\\
\G_i\ast t(h)&=\G_{i-1}\ast\left[\frac{1}{g(s_{i-1},h)}\g_{\D_{i-1}-\D_i}\star g(s_i,h)^{\frac{s_{i-1}}{s_i}}\right]\ \ \ \ \ \ \ \ i=2,\ldots,k
\end{split}
\label{ricorsione_Gtilde}
\eeq
and $t(h)$ is a test function.
Note that the first line of Eq. (\ref{derivative_int}), namely the term with the derivative with respect to $\Delta^f$ is due to the fact that when we obtain that relation we are deriving Eq. (\ref{interaction_zero}) with respect to the $\Delta_i$. In the representation (\ref{interaction_zero}) we have an extra dependence of $\Delta^r$ as a function of $\Delta_1$ and this is the reason why we get the derivative with respect to $\Delta^f$.
We now define
\beq
\tilde \G_i\equiv\int_{-\infty}^\infty \DD \z \left[\g_{\Delta_\g(\z)}\star \left(\eee^{h+\eta-(\D_\g(\z)+\D_1)/2}\Theta^m\left[\frac{h+\eta-(\D_\g(\z)+\D_1-\D^g)/2}{\sqrt{2\D^g}}\right]\right)\right]\G_i
\eeq
so that
\beq
\begin{split}
\tilde \G_1\ast t(h)&=\int_{-\infty}^\infty \DD \z \left[\g_{\Delta_\g(\z)}\star \left(\eee^{h+\eta-(\D_\g(\z)+\D_1)/2}\Theta^m\left[\frac{h+\eta-(\D_\g(\z)+\D_1-\D^g)/2}{\sqrt{2\D^g}}\right]\right)\right]\eee^{sf(s_1,h)}t(h)\\
\tilde \G_i\ast t(h)&=\tilde\G_{i-1}\ast\left[\frac{1}{g(s_{i-1},h)}\g_{\D_{i-1}-\D_i}\star g(s_i,h)^{\frac{s_{i-1}}{s_i}}\right]\ \ \ \ \ \ \ \ i=2,\ldots,k
\end{split}
\eeq
Then Eq. (\ref{derivative_int}) can be rewritten as
\beq
\begin{split}
\frac 2d \frac{\partial s_{int}}{\partial G_i}=\ &\frac{\wh \f_g}{2} \int_{-\infty}^\infty \de h P(s_1,h)\left[ \left(s f'(s_1,h)\right)^2+sf''(s_1,h)+sf'(s_1,h)    \right]\left(\frac{1}{s_1}\delta_{i,1}+(1-\delta_{i,1})\left(\frac{1}{s_i}-\frac{1}{s_{i-1}}\right)\right)\\
&+s\frac{\wh \f_g}{2}\int_{-\infty}^\infty\de h \left\{\frac{s_i-s_{i-1}}{s_i}\tilde \G_i \ast \left(f'(s_i,h)\right)^2+\left(\frac{1}{s_i}-\frac{1}{s_{i-1}}\right)\sum_{j=1}^{i-1}(s_j-s_{j-1})\tilde \G_j\ast\left(f'(s_j,h)\right)^2\right\}\:.
\end{split}
\eeq
At this point we introduce $P(s_i,h)$ that are defined in the following way
\beq
\int_{-\infty}^\infty\de h \tilde \G_i\ast t(h)=\int_{-\infty}^\infty \de h P(s_i,h) t(h)
\eeq
where $t(h)$ is a test function.
Using Eq. (\ref{ricorsione_Gtilde}) we get \cite{CKPUZ14JSTAT} the following recursion relation for $P$
\beq
\begin{split}
P(s_1,h)&=\eee^{h+\eta-\Delta_1/2}\int_{-\infty}^\infty \DD \z \int_{-\infty}^\infty\frac{\de x}{\sqrt{2\pi \Delta_\g(\z)}}\eee^{-\frac{1}{2\Delta_\g(\z)}\left(x+\frac{\Delta_\g(\z)}{2}\right)^2} \Theta^{m}\left[\frac{h+\eta -x+(\Delta^g-\Delta_1)/2}{\sqrt{2\Delta^g}}\right]\eee^{s f(s_1,h-x+\Delta_\g(\z)/2)}\\
P(s_i,h)&=\int_{-\infty}^\infty \de z \frac{P(s_{i-1},z)}{g(s_{i-1},z)}\g_{\Delta_{i-1}-\Delta_i}(h-z)g(s_i,h)^{s_{i-1}/s_i}
\end{split}
\eeq
Then the derivative of the interaction term with respect to $G_i$ is given by
\beq
\begin{split}
\frac 2d \frac{\partial s_{int}}{\partial G_i}=\ &\frac{\wh \f_g}{2} \int_{-\infty}^\infty \de h P(s_1,h)\left[ \left(s f'(s_1,h)\right)^2+sf''(s_1,h)+sf'(s_1,h)    \right]\left(\frac{1}{s_1}\delta_{i,1}+(1-\delta_{i,1})\left(\frac{1}{s_i}-\frac{1}{s_{i-1}}\right)\right)\\
&+s\frac{\wh \f_g}{2}\int_{-\infty}^\infty\de h \left\{\frac{s_i-s_{i-1}}{s_i}P(s_i,h) \left(f'(s_i,h)\right)^2+\left(\frac{1}{s_i}-\frac{1}{s_{i-1}}\right)\sum_{j=1}^{i-1}(s_j-s_{j-1})P(s_j,h)\left(f'(s_j,h)\right)^2\right\}
\end{split}
\eeq
{Collecting all these results, we} finally {obtain} the variational equation for $G_i$:
\beq
\begin{split}
&\delta_{i,1}\frac{m}{m G_1+ms\Delta^f+s\Delta^g}+\left(\frac{s}{s_i}-\frac{s}{s_{i-1}}\right)\frac{1}{G_i}\\
&=-\frac{\wh \f_g}{2} \int_{-\infty}^\infty \de h P(s_1,h)\left[ \left(s f'(s_1,h)\right)^2+sf''(s_1,h)+sf'(s_1,h)    \right]\left(\frac{1}{s_1}\delta_{i,1}+(1-\delta_{i,1})\left(\frac{1}{s_i}-\frac{1}{s_{i-1}}\right)\right)\\
&-s\frac{\wh \f_g}{2}\int_{-\infty}^\infty\de h \left\{\frac{s_i-s_{i-1}}{s_i}P(s_i,h) \left(f'(s_i,h)\right)^2+\left(\frac{1}{s_i}-\frac{1}{s_{i-1}}\right)\sum_{j=1}^{i-1}(s_j-s_{j-1})P(s_j,h)\left(f'(s_j,h)\right)^2\right\}
\end{split}
\eeq
It is very useful to {consider separately the cases} $i=1$ and $i>1$. In the first case we have the following equation
\beq
\begin{split}
\frac{m}{m G_1+ms\Delta^f+s\Delta^g}+\left(\frac{s}{s_1}-1\right)\frac{1}{G_1}=&-s\frac{\wh \f_g}{2}\int_{-\infty}^\infty\de h \frac{s_1-s}{s_1}P(s_1,h) \left(f'(s_1,h)\right)^2\\
&-\frac{\wh \f_g}{2s_1} \int_{-\infty}^\infty \de h P(s_1,h)\left[ \left(s f'(s_1,h)\right)^2+sf''(s_1,h)+sf'(s_1,h) \right]\:,
\end{split}
\eeq
and using Eq. (\ref{equazione_Delta_f_finite_s}) we get
\beq
\frac{m\Delta^f+\Delta^g}{G_1(mG_1+s\Delta^f+s\Delta^g)}=\frac{\wh\f_g}{2}\int_{-\infty}^\infty\de h P(s_1,h)\left(f'(s_1,h)\right)^2,
\eeq
that in the continuum limit and for $s\to 0$ becomes
\beq
\frac{m\Delta^f+\Delta^g}{mG(0)^2}=\frac{\wh \f_g }{2}\int_{-\infty}^\infty\de h\, P(0,h) \left(f'(0,h)\right)^2\:.
\eeq
Let us now consider the case $i>1$. We get
\beq
\begin{split}
&\left(\frac{1}{s_i}-\frac{1}{s_{i-1}}\right)\frac{1}{G_i}=-\left(\frac{1}{s_i}-\frac{1}{s_{i-1}}\right)\frac{\wh \f_g}{2} \int_{-\infty}^\infty \de h P(s_1,h)\left[ s(f'(s_1,h))^2+f''(s_1,h)+f'(s_1,h) \right]\\
&-\frac{\wh \f_g}{2}\int_{-\infty}^\infty\de h \left\{\frac{s_i-s_{i-1}}{s_i}P(s_i,h) \left(f'(s_i,h)\right)^2+\left(\frac{1}{s_i}-\frac{1}{s_{i-1}}\right)\sum_{j=1}^{i-1}(s_j-s_{j-1})P(s_j,h)\left(f'(s_j,h)\right)^2\right\},
\end{split}
\eeq
{and again we can use} Eq. (\ref{equazione_Delta_f_finite_s}) to get
\beq
\begin{split}
&\left(\frac{1}{s_i}-\frac{1}{s_{i-1}}\right)\frac{1}{G_i}=\left(\frac{1}{s_i}-\frac{1}{s_{i-1}}\right)\frac{m}{mG_1+ms\Delta^f+s\Delta^g }\\
&-\frac{\wh \f_g}{2}\int_{-\infty}^\infty\de h \left\{\frac{s_i-s_{i-1}}{s_i}P(s_i,h) \left(f'(s_i,h)\right)^2+\left(\frac{1}{s_i}-\frac{1}{s_{i-1}}\right)\sum_{j=1}^{i-1}(s_j-s_{j-1})P(s_j,h)\left(f'(s_j,h)\right)^2\right\}
\end{split}
\eeq
which can be easily rewritten in the limit $s\to 0 $ as
\beq
\frac{1}{G_i}=\frac{1}{G_1}+\frac{\wh \f_g}{2}\int_{-\infty}^\infty\de h\, \left\{s_{i-1} P(s_i,h)\left(f'(s_i,h)\right)^2-\sum_{j=1}^{i-1}(s_j-s_{j-1})P(s_j,h)\left(f'(s_j,h)\right)^2\right\};
\eeq
and in the continuum limit we get for $x>0$
\beq
\frac{1}{G(x)}=\frac{1}{G(0)}+\frac{\wh \f_g}{2} \int_{-\infty}^\infty\de h x\left\{x P(x,h)\left(f'(x,h)\right)^2-\int_{0}^x\de y P(y,h)\left(f'(y,h)\right)^2\right\}
\label{equazione_Gx}
\eeq
Note that we can also safely {take the limit $x\to0$ of} this equation, proving the continuity of the solution of the fullRSB equations.
Here we summarize the finite $k$ RSB equations for $s\to 0$:
\beq
\begin{split}
f(1,h)&=\log \g_{\Delta_k}\star \theta(h)=\log \Theta\left[\frac{h}{\sqrt{2\Delta_k}}\right]\\
f(s_i,h)&=\frac{1}{s_i}\log \g_{\Delta_i-\Delta{i+1}}\star \eee^{s_i f(s_{i+1},h)}\\
P(s_1,h)&=\eee^{h+\eta-\Delta_1/2}\int_{-\infty}^\infty \DD \z \int_{-\infty}^\infty\frac{\de x}{\sqrt{2\pi \Delta_\g(\z)}}\eee^{-\frac{1}{2\Delta_\g(\z)}\left(x+\frac{\Delta_\g(\z)}{2}\right)^2} \Theta^{m}\left[\frac{h+\eta -x+(\Delta^g-\Delta_1)/2}{\sqrt{2\Delta^g}}\right]\\
P(s_i,h)&=\int_{-\infty}^\infty \de z \frac{P(s_{i-1},z)}{g(s_{i-1},z)}\g_{\Delta_{i-1}-\Delta_i}(h-z)g(s_i,h)^{s_{i-1}/s_i}\\
\frac{1}{G_1}&=-\frac{\wh \f_g}{2}\int_{-\infty}^\infty \de h P(s_1,h)\left(f''(s_1,h)+f'(s_1,h)\right)\\
\frac{m\Delta^f+\Delta^g}{mG_1^2}&=\frac{\wh \f_g}{2}\int_{-\infty}^\infty\de h\, P(s_1,h) \left(f'(s_1,h)\right)^2\\
\k_i&=\frac{\wh \f_g}{2}\int_{-\infty}^\infty \de h P(s_i,h)\left(f'(s_i,h)\right)^2\\
\frac{1}{G_i}&=\frac{1}{G_1} +s_{i-1}\k_{i}-\sum_{j=1}^{i-1}(s_i-s_{i-1})\k_j\ \ \ \ \ i=2,\ldots, k
\end{split}
\eeq
These equations can be easily solved {numerically by iterations}.
Moreover in the continuum $k\to \infty$ limit we get
\beq
\begin{split}
f(1,h)&=\log \g_{\Delta(1)}\star \theta(h)=\log \Theta\left[\frac{h}{\sqrt{2\Delta(1)}}\right]\\
\frac{\partial f}{\partial x}&=\frac 12\frac{\dot G(x)}{x}\left[\frac{\partial^2 f}{\partial h^2}+x\left(\frac{\partial f}{\partial h}\right)^2\right]\\
P(0,h)&=\eee^{h+\eta-\Delta(0)/2}\int_{-\infty}^\infty \DD \z \int_{-\infty}^\infty\frac{\de x}{\sqrt{2\pi \Delta_\g(\z)}}\eee^{-\frac{1}{2\Delta_\g(\z)}\left(x+\frac{\Delta_\g(\z)}{2}\right)^2} \Theta^{m}\left[\frac{h+\eta -x+(\Delta^g-\Delta(0))/2}{\sqrt{2\Delta^g}}\right]\\
\dot{P}(x,h)&= -\frac{\dot{G}(x)}{2x}\left[P''(x,h)-2x(P(x,h)f'(x,h))'\right]\\
\frac{1}{G(0)}&=-\frac{\wh \f_g}{2}\int_{-\infty}^\infty \de h P(0,h)\left(f''(0,h)+f'(0,h)\right)\\
\frac{m\Delta^f+\Delta^g}{mG(0)^2}&=\frac{\wh \f_g}{2}\int_{-\infty}^\infty\de h\, P(0,h) \left(f'(0,h)\right)^2\\
\k(x)&=\frac{\wh \f_g}{2}\int_{-\infty}^\infty \de h P(x,h)\left(f'(x,h)\right)^2\\
\frac{1}{G(x)}&=\frac{1}{G(0)} +x\k(x)-\int_{0}^x\de y\k(y)\ \ \ \ \ \ \ \ \ \ \ \ x>0
\label{fullRSB_continue}
\end{split}
\eeq
At this point we can show that the two equations (\ref{equazione_G0}) and (\ref{equazione_Gx}) can be rewritten in a single equation that is valid for $x\in [0,1]$
\beq
\frac{1}{G(x)}=-\frac{\wh \f_g}{2}\int_{-\infty}^\infty \de h \left(P(x,h)f''(x,h)+P(0,h)f'(0,h)\right)
\label{alternativa}
\eeq
This can be easily shown by taking the derivative of Eq. (\ref{equazione_Gx}) with respect to $x$ and using the equations (\ref{fullRSB_continue}) \cite{CKPUZ14JSTAT}.
Finally we can consider the derivatives with respect to $x$ of equation (\ref{alternativa}). For $x$ that is such that $\dot G(x)\neq 0$ we get
\beq
1=\frac{\wh \f_g}{2}\int_{-\infty}^\infty\de h P(x,h)\left(G(x)f''(x,h)\right)^2
\label{MG_SF}
\eeq
This equation coincides with the one found in \cite{CKPUZ14JSTAT}.


\subsection{Lagrange multipliers}
A different route to obtain the fullRSB equations is to start directly from the $k\to\infty$ limit, and take the functional derivatives of the  $s_g$ with respect to $\D(x)$, introducing Lagrange multipliers to enforce Eq. (\ref{Parisi_eq}) and its initial condition \cite{SD84}. These Lagrange multipliers are called $P(x,h)$ and $P(1,h)$ and we will see in this section that they are the exact same functions we defined in (\ref{fullRSB_continue}).
In order to obtain the fullRSB equations, we rewrite the relevant part of the free energy of the followed system \eqref{eq:sfull} (we omit constant terms), adding the Lagrange multipliers
\beq
\begin{split}
\frac 2d \mathcal{S}_\infty=&-s\int_s^1\frac{dx}{x^2}\log G(x) + \log\left[mG(s)+ms\Delta^f+s\Delta^g\right]\\
&-\wh\varphi_g\int_{-\infty}^\infty\DD\z \int_{-\infty}^{\infty} dh \ e^{h}\left\{1-\Th\left(  \frac{h + \D^g/2}{\sqrt{2\D^g}} \right)^m \g_{\Delta_\g(\z)}\star \eee^{sf(s,h-\eta+(\Delta_\g(\z)+\Delta(s))/2}\right\}\\
&+s\wh\varphi_g\int_s^1 dx\ \int_{-\infty}^{\infty}dh\ P(x,h)\left\{\dot{f}(x,h)-\frac{\dot{G}(x)}{2x}\left[f''(x,h)+xf'(x,h)^2\right]\right\}\\
&-s\wh\varphi_g\int_{-\infty}^{\infty}dh\ P(1,h)\left\{f(1,h)-\log\Th\left(\frac{h}{\sqrt{2G(1)}}\right)\right\}\:.
\end{split}
\eeq
We start by taking the derivative with respect to $\D_f$. We get
\beq
\begin{split}
\frac{ms}{mG(s)+ms\Delta^f+s\Delta^g}=&-\frac{\wh\f_g}{2}\int_{-\infty}^\infty\DD \z\int_{-\infty}^\infty \de h \left[\g_{\Delta_\g(\z)}\star \eee^{h+\eta-(\Delta^f+\Delta(s)-\Delta^g)/2}\Theta\left[\frac{h+\eta-(\Delta^f+\Delta(s)-\Delta^g)/2}{\sqrt{2\Delta^g}}\right]^m\right]\\
&\times \left[\left(sf'(s,h)\right)^2+sf'(s,h)+sf''(s,h)\right]\eee^{sf(s,h)},
\label{eq_Delta_f_s_finito}
\end{split}
\eeq
that coincides with the continuum limit of Eq. (\ref{equazione_Delta_f_finite_s})\footnote{In order to perform the calculation we have only used the differential representation of the convolution of a Gaussian function (\ref{differential_gaussian}) and an integration by parts. This is analogous to what has been done {to derive the} (\ref{interazione_Delta_f}).}.
Varying $S_{\infty}$ with respect to $P(x,h)$ and $f(x,h)$ we get
\begin{eqnarray}
\dot{f}(x,h)&=&\frac{\dot{G}(x)}{2x}\left[f''(x,h)+xf'(x,h)^2\right],\label{eq:eqf}\\
\dot{P}(x,h)&=& -\frac{\dot{G}(x)}{2x}\left[P''(x,h)-2x(P(x,h)f'(x,h))'\label{eq:eqP}\right].
\end{eqnarray}
By differentiating with respect to $f(s,h)$ we get
\beq
\begin{split}
P(s,h)&= e^{h+\eta-\D(s)/2}\int_{-\infty}^\infty\DD \z \int_{-\infty}^{\infty} dx \frac{e^{-\frac{(x+\D_\g(\z)/2)^2}{2\D_\g(\z)}}}{\sqrt{2\pi\D_\g(\z)}}\Th^m\left(\frac{h-x+\eta-\D(s)/2+\D^g/2}{\sqrt{2\D^g}}\right)\eee^{sf(s,h)}
\end{split}
\label{eq:initialP}
\eeq
that coincides with (Eq. \ref{defP1}).
We now differentiate with respect to $G(x)$. We get
\beq
\begin{split}
0=&\ \delta(x-s)\frac{m}{mG(s)+ms\Delta^f+s\Delta^g}-\frac{s}{x^2}\frac{1}{G(x)}\\
&\ +\wh\f_g\frac{\delta \Delta(s)}{\delta G(x)}\frac{\delta}{\delta \Delta(s)}\int_{-\infty}^\infty\DD\z \int_{-\infty}^\infty\de h \eee^h\Theta^m\left[\frac{h+\Delta^g/2}{\sqrt{2\Delta^g}}\right]\g_{\Delta_\g(\z)}\star \eee^{sf(s,h-\eta+(\Delta^f+\Delta(s))/2)}\\
&\ -\frac{s\wh \f_g}{2}\int_{-\infty}^\infty\de h \left[\delta(x-1)P(1,h)\left(f''(1,h)+(f'(1,h))^2\right)-\frac 1s\delta(x-s)P(s,h)\left(f''(s,h)+s(f'(s,h))^2\right)\right]\\
&\ +\frac{s\wh \f_g}{2}\int_{-\infty}^\infty \de h \frac{\de}{\de x}\left[\frac 1x P(x,h)\left(f''(x,h)+x\left(f'(x,h)\right)^2\right)\right]+\wh \f_g s \delta(x-1)\int_{-\infty}^\infty\de h P(1,h)\frac{\delta }{\delta \Delta(1)}\log \Theta\left[\frac{h}{\sqrt{2\Delta(1)}}\right]
\end{split}
\eeq
{Now,} first we note that the terms proportional to $\delta(x-1)$ simplify because of the initial condition for $f(x,h)$.
Moreover we have that
\beq
\frac{s\wh\f_g}{2}\int_{-\infty}^\infty \de h \frac{\de}{\de x}\left[\frac 1x P(x,h)\left(f''(x,h)+x\left(f'(x,h)\right)^2\right)\right]=-\frac{1}{x^2}\int_{-\infty}^\infty\de h P(x,h)f''(x,h)
\eeq
and
\beq
\begin{split}
&\wh\f_g\frac{\delta \Delta(s)}{\delta G(x)}\frac{\delta}{\delta \Delta(s)}\int_{-\infty}^\infty\DD\z \int_{-\infty}^\infty\de h \eee^h\Theta^m\left[\frac{h+\Delta^g/2}{\sqrt{2\Delta^g}}\right]\g_{\Delta_\d(\z)}\star \eee^{sf(s,h-\eta+(\Delta^f+\Delta(s))/2)}=\\
&\frac{\wh \f_g}{2}\left(\delta(x-s)-\frac{s}{x^2}\right)\int_{-\infty}^\infty\de h P(s,h)f'(s,h)\:.
\end{split}
\eeq
Putting all the pieces together we get
\beq
\begin{split}
0&=\delta(x-s)\left[\frac{m}{mG(s)+ms \Delta^f+s\Delta^g}+\frac{\wh\f_g}{2}\int_{-\infty}^\infty\de h P(s,h)\left[s\left(f'(s,h)\right)^2+f'(s,h)+f''(s,h)\right]\right]\\
&-\frac{s}{x^2}\left[\frac{1}{G(x)}+\frac{\wh \f_g}{2}\int_{-\infty}^\infty\de h \left(P(x,h)f''(x,h)+P(s,h)f'(s,h)\right)\right]
\end{split}
\eeq
and using Eq. (\ref{eq_Delta_f_s_finito}) we get
\beq
\frac{1}{G(x)}=-\frac{\wh \f_g}{2}\int_{-\infty}^\infty\de h \left(P(x,h)f''(x,h)+P(s,h)f'(s,h)\right) \ \ \ \ \ \ \ \ \ \ x\in[s,1]\:.
\eeq
Looking at $x=s$ and using again Eq. (\ref{eq_Delta_f_s_finito}) we obtain
\beq
\frac{m\Delta^f+\Delta^g}{G(s)\left(mG(s)+ms \Delta^f+\Delta^g\right)}=\frac{\wh \f_g}{2}\int_{-\infty}^\infty\de h P(s,h)\left(f'(s,h)\right)^2
\eeq
that, in the $s\to 0$ limit become Eq.s (\ref{fullRSB_continue}) and (\ref{alternativa}).
This completes the proof that the equations we get using the Lagrange multipliers are the same as the ones we have obtained from the finite $k$RSB approach.
It can be easily checked that once a 1RSB parametrization $\Delta(x)=\Delta$ is chosen we get back the saddle point equations of \cite{RUYZ15}.

\subsection{The fullRSB equations at equilibrium ($m=1$)}
We have seen in section \ref{sec:m=1} that the replicated entropy simplifies once we consider a planted equilibrium configuration, namely once the number of replicas in the master system $m$ goes to 1. This case is important not only because it corresponds to a physical working assumption but because it gives much more simple saddle point equations. 
In this case the fullRSB equations in the continuum limit are exactly the same as Eq.s (\ref{fullRSB_continue}) with the only difference that the initial condition for $P(0,h)$ changes and it is given by
\beq
P(0,h) = e^{h+\eta-\D(0)/2}\int_{-\infty}^\infty d\z\ \frac{e^{-\frac{\zeta^2}{2}}}{\sqrt{2\pi}}\Th\left(\frac{h+\eta + \D_r+\zeta^2\gamma^2/2-\D(0)}{\sqrt{2(2\D_r+\zeta^2\gamma^2-\D(0))}}\right)\:.
\eeq
Moreover it is quite simple to see that the equation that determines $\Delta^f$ becomes an equation for $\Delta^r$ that is
\beq
\frac{2\Delta^r-\Delta(0)}{G(0)^2}=\frac{\wh \f_g}{2}\int_{-\infty}^\infty\de h\, P(0,h) \left(f'(0,h)\right)^2\:.
\eeq

\section{Physical observables}

{Now that the entropy $s_g$ of the glass state has been computed,} we {can investigate} the response of a glass state once a perturbation is applied. We have developed the formalism to compute two kind of responses \cite{RUYZ15}: the first one is the pressure, that is the response of the system to a small variation in the diameter of the spheres; the second one is the, shear stress that is the response of the glass once a shear strain is applied. 
{We note that our formalism allow us to obtain only the average of the responses over all the glass states that are contained in a typical metastable basin.}

\subsection{Pressure}
Let us start with the reduced pressure. It is defined as \cite{Hansen}
$$
p  \equiv \frac{\beta P}{\rho} = -\frac{ds_g}{d\eta} = -\frac{\partial s_g}{\partial \eta}.
$$
The dependence of $s_g$ from $\eta$ is all contained in the $P(0,h)$ function in the interaction term; {however, we can shift it to $f(h,0)$ by translating $h$ in the integral. Then} taking the derivative we get
\beq
p = \frac{d\wh\varphi_g}{2}\int_{-\infty}^\infty dh\ P(0,h)f'(0,h).
\label{eq:pressureP0}
\eeq
{This result is quite clear: the pressure is given in terms of quantities computed at $x=0$ which means that we have integrated out the whole fullRSB hierarchy of marginal states to get the full averaged response. However let us consider
\beq
\frac{d}{dx}\int_{-\infty}^\infty dh\ P(x,h)f'(x,h) = \int_{-\infty}^\infty dh\ [\dot{P}(x,h)f'(x,h) + P(x,h)\dot{f}'(x,h)].
\eeq
We can expand this expression using the equations of motion \eqref{fullRSB_continue}. Using the symbol $a \sim b$ to denote that $\int_{-\infty}^\infty a=\int_{-\infty}^\infty b$, we get
\beq
\begin{split}
\dot{P}f' + P\dot{f}' =\  &-\frac{\dot{G}}{2x}[P''-2x(Pf')']f'+\frac{\dot{G}}{2x}[f'''+2xf'f'']P\\
\sim\ & -\frac{\dot{G}}{2x}Pf''' -\dot{G}Pf'f''+\frac{\dot{G}}{2x}Pf''' + \dot{G}Pf'f'' = 0
\end{split}
\eeq
So we get that the expression for the pressure \eqref{eq:pressureP0} is actually independent of $x$. In particular we can write
\beq
p = \frac{d\wh\varphi_g}{2}\int_{-\infty}^\infty dh\ P(1,h)f'(1,h).
\label{eq:pressureP1}
\eeq
which gives the pressure in terms of quantities that are directly related to fullRSB microstates.
This means that the pressure in different microstates does not fluctuate.\\
}
%
%
\begin{figure}
\centering
\includegraphics[scale=1]{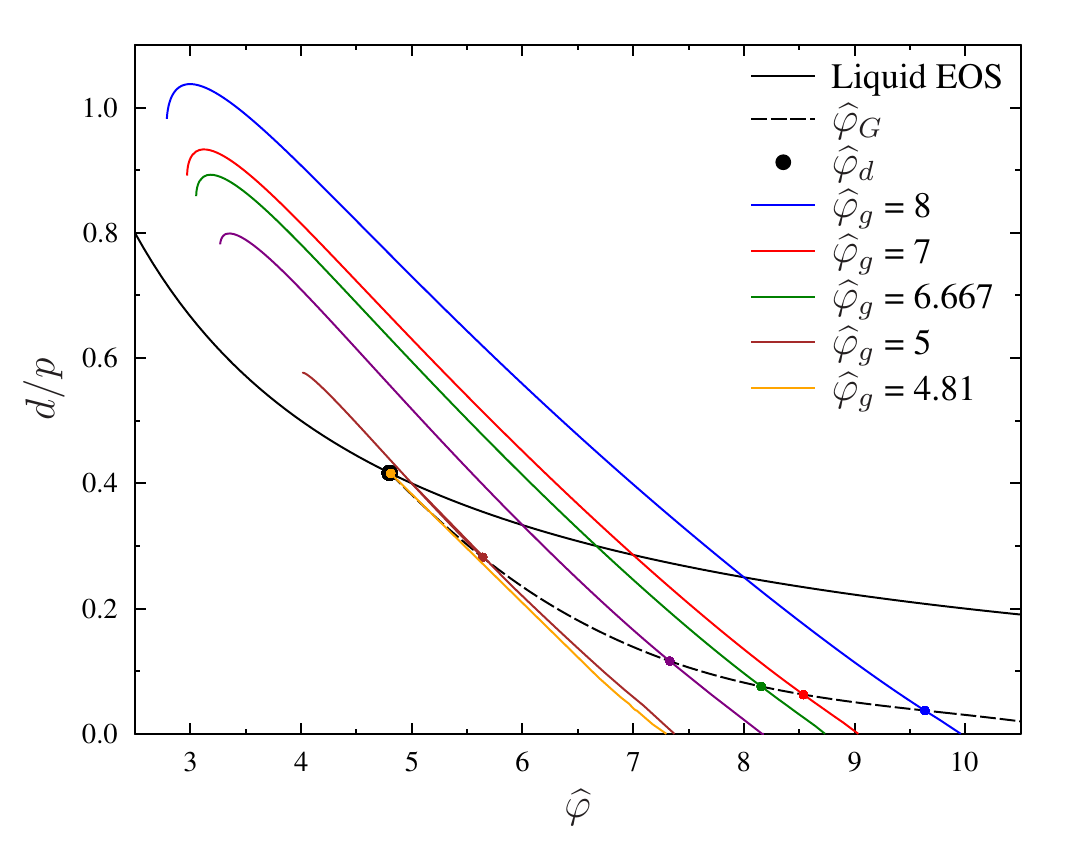}
\caption{{Equations of state of various glasses at different planting densities $\wh \f_g$ (i.e. the inverse reduced pressure $d/p$ as a function of the reduced packing fraction $\wh \f =2^d\f/d$). The black line represents the equation of state $d/p=1/\wh \f$ of the liquid. Beyond the dynamical point $\wh \f_d$ the system develops an exponential number of metastable states, wherein the system can get trapped and start to age, forming a glass. The colored lines represent the equation of state of different glasses, each one planted at a given packing fraction $\wh \f_g$}. The intersection between these lines and the dashed line gives the Gardner transition point $\wh \f_G$ for each one of these glasses; beyond that point we show the equation of states obtained solving the fullRSB equations.
}
\label{SF_compression}
\end{figure}
{In Fig. \ref{SF_compression} we show the result obtained by plotting the inverse rescaled pressure as a function of $\wh \f = \wh \f_g e^\eta$, for various planting densities as is \cite{RUYZ15}. One can see that, differently from the study of \cite{RUYZ15}, we can now follow the states also in the Gardner phase, all the way to the jamming threshold. We estimate the jamming density $\wh \f_j$ of the states planted at $\wh \f_g = \wh \f_d$ and we obtain.}
\beq
\wh \f_j(\wh \f_d) \simeq 7.30
\eeq
This value corresponds to the rescaled jamming density of the densest packing that can be constructed via an annealing-like procedure on timescales that are polynomial in the system size in the infinite dimensional limit. Less dense packings can in principle be constructed, but the computation of their jamming density must be performed using the formalism of \cite{CKPUZ14JSTAT}. 
We remark that this result could not be obtained, even approximately, with the RS ansatz since in that case unphysical spinodal points made it impossible to follow down to jamming the less dense states planted near the dynamical transition, as discussed in \cite{RUYZ15} and in analogy with similar results for spin glasses \cite{KZ10,KZ10b,SCKLZ12}. However, at least in the case of hard spheres, the fullRSB ansatz alone is able to cure these artifacts, with no need to generalize the FP construction to a chain of three or more replicas \cite{FP13,FPU15} as it was suggested in \cite{KZ10,SCKLZ12}.

\subsection{Shear stress}
\begin{figure}
\centering
\includegraphics[scale=1]{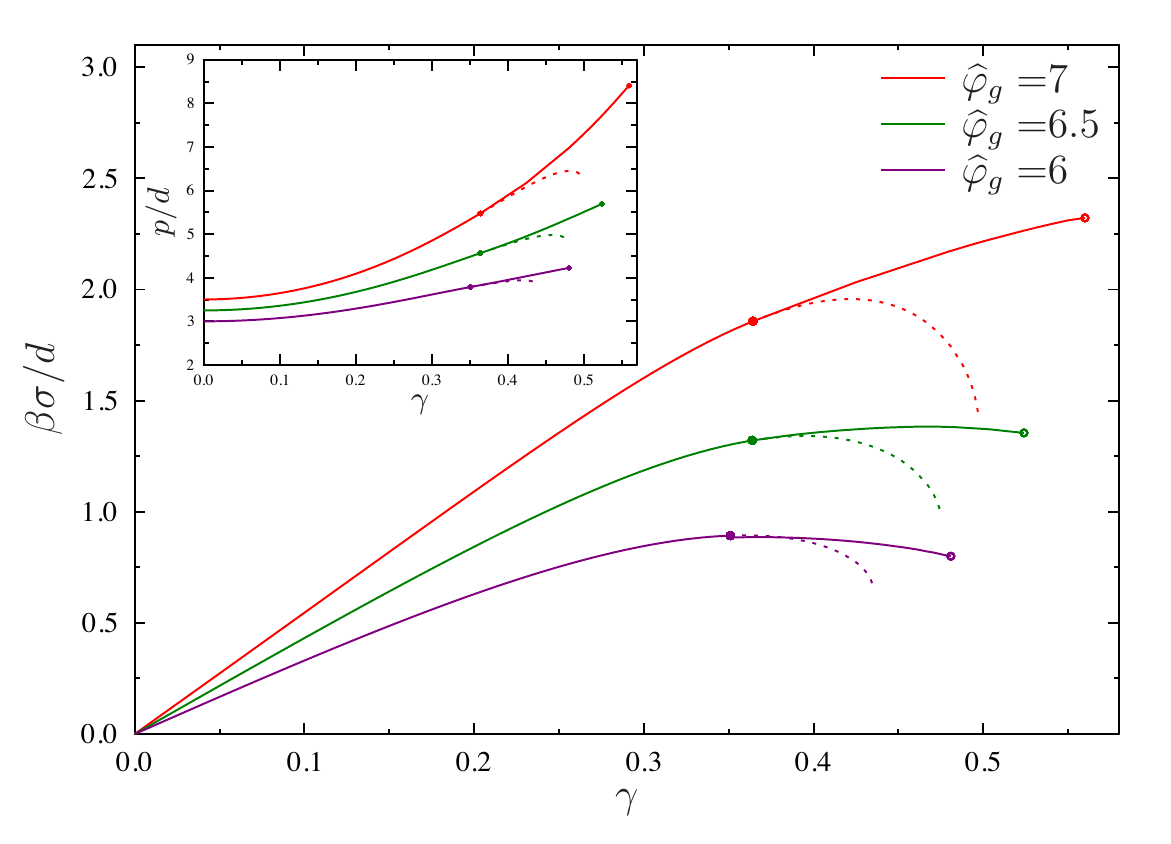}
\caption{The behavior of the shear stress $\s$ as a function of the shear strain $\g$ for three glasses planted at three different packing fractions.  
In the inset we also plot the behavior of the pressure during the evolution of the glassy states under strain; again as in \cite{RUYZ15}, dilatancy is observed \cite{Ti14}. The full circles along state curves represent the Gardner transition point for the three glasses, {while the empty ones represent the endpoint beyond which we find hard to solve numerically the fullRSB equations.} 
}
\label{fig_shear}
\end{figure}

Let us now compute the shear stress that is defined as
\beq
\frac{ds_g}{d\gamma} = \frac{\partial s_g}{\partial \gamma}.
\eeq
The dependence of $s_g$ on $\gamma$ is all in the $P(0,h)$, so we can easily get at equilibrium
\beq
\s = \gamma\frac{d\wh\varphi_g}{2}\int dh\ e^{h-\D(0)/2}\int d\zeta\ \frac{e^{-\frac{\zeta^2}{2}}}{\sqrt{2\pi}}\ \zeta^2\ \frac{e^{-\frac{(h + \D_r+\zeta^2\gamma^2/2-\D(0))^2}{2(2\D_r+\zeta^2\gamma^2-\D(0))}}}{\sqrt{2\pi(2\D_r+\zeta^2\gamma^2-\D(0))}}\left(\frac{\D_r+\zeta^2\gamma^2/2-h}{2\D_r+\zeta^2\gamma^2-\D(0)}\right)f(0,h)\:.
\eeq
In Fig. \ref{fig_shear} we show the behavior of the shear stress as a function of the shear strain for three different glassy states planted at three different packing fractions. {Again, we are able to follow the states beyond the Gardner point and a stress overshoot is detected as in \cite{RUYZ15}, in analogy with numerical and experimental observations \cite{RTV11,Kou12}. However, we are unable to follow the states all the way to the yielding point $\g_Y$ whereupon the glass would yield and start to flow \cite{RTV11}}, since the numerical solution of the fullRSB equations is extremely hard.

\section{Scaling analysis near jamming}
We show now that once the glass state is followed in compression up to the jamming point, the solution on the fullRSB equations (\ref{fullRSB_continue}) develops a scaling regime characterized by a set of critical exponents that coincides with the ones computed in \cite{CKPUZ14JSTAT}. The proof will be given by showing that the scaling equations close to jamming, and the asymptotic behavior of their initial conditions are the same as those that had been obtained in \cite{CKPUZ14JSTAT}; the {values of the} critical exponents follow directly from these requirements.

\subsection{Scaling form of the equations}
On approaching the jamming point, the mean square displacement {of the fullRSB microstates} goes to zero. We thus define the jamming limit as $\Delta(1)\equiv \Delta_{EA}\to 0$ \cite{PZ10}. Moreover we expect that $\Delta^f$ stays finite.\\
We want to show that the fullRSB equations develop a scaling regime. At jamming the pressure diverges as $1/p\propto\D_{EA}^{1/\k}$ \cite{IBB12,CKPUZ14JSTAT} and we want to determine $\k$. We thereby define the following scaling variables and functions:
\begin{eqnarray}
 y &\equiv&\D_{EA}^{-\frac{1}{\kappa}} x,\\
 \wh{f}(y,h) &\equiv& \scl f(\scl y,h),\\
 \g(y) &\equiv& \frac{G(\scl y)}{\scl},\\
 \wh{P}(y,h) &\equiv& e^{-h-\eta}P(\scl y,h)\:.
\end{eqnarray}
The initial conditions for the new functions $\wh{f}$ and $\wh{P}$ are therefore
\begin{eqnarray}
\wh{P}(0,h) &=& e^{-\D(0)/2}\int_{-\infty}^{\infty} dx \frac{e^{-\frac{(x+\D_f)^2}{2\D_f}}}{\sqrt{2\pi\D_f}}\Th\left(\frac{h-x+\eta-\D(0)/2+\D_g/2}{\sqrt{2\D_g}}\right)^m,\\
\wh{f}(1/\scl,h) &=& \scl\log\Theta\left(\frac{h}{\sqrt{2\scl\gamma(1/\scl)}}\right),
\end{eqnarray}
and the relation between $\D(y)$ and $\gamma(y)$ becomes
\beq
\D(y)=\frac{\gamma(y)}{y}-\int_{y}^{1/\scl}\frac{dz}{z^2}\gamma(z).
\label{eq:deltaofgamma}
\eeq
{Using the same reasoning,} the variational equations for the scaling functions are:
\begin{eqnarray*}
\frac{\partial\wh{f}(y,h)}{\partial y}&=&\frac{\dot{\g}(y)}{2y}\left[\frac{\partial^2\wh{f}(y,h)}{\partial h^2}+y\left(\frac{\partial\wh{f}(y,h)}{\partial h}\right)^2\right],\\
\frac{\partial\wh{P}(y,h)}{\partial y}&=& -e^{-h}\frac{\dot{\g}(y)}{2y}\left[\frac{\partial^2[e^h\wh{P}(y,h)]}{\partial h^2}-2y\frac{\partial}{\partial h}\left(e^h\wh{P}(y,h)\frac{\partial \wh{f}(y,h)}{\partial h}\right)\right]\\
\frac{1}{\gamma(y)} &=&\frac{1}{\la\D\ra} + y\kappa(y)-\int_0^y dz\ \kappa(z) \\
\frac{m\D_f+\D_g}{m\la\D\ra^2} &=& \kappa(0)\\
\kappa(y) &=& \frac{\wh \varphi_g e^\eta}{2} \int_{-\infty}^{\infty} dh\ e^h \wh{P}(y,h)\wh{f}'(y,h)^2,
\label{scaling_eqs}
\end{eqnarray*}
{which} are very close to those obtained in \cite{CKPUZ14JSTAT}. {The entropy, for its part, is rephrased in}  
\beq
\begin{split}
\mathcal{S} = &\frac{1}{2}\log\left(\frac{\pi\la\D\ra}{d^2}\right)+\frac{1}{2}\log(\scl)\\
&+ \frac{1}{\scl}\left[-\frac{1}{2}\int_0^{1/\scl}\frac{dy}{y^2}\log\left(\frac{\gamma(y)}{\la\D\ra}\right) + \frac{1}{2}\frac{m\D_f+\D_g}{m\la\D \ra} + \frac{\wh\varphi_g e^\eta}{2}\int_{-\infty}^{\infty} dh\ e^h\wh{P}(0,h)\wh{f}(0,h)\right],
\label{eq:sclentropy}
\end{split}
\eeq
where now $\la\D\ra$ is defined as
\beq
\la\D\ra = \int_0^{1/\scl}dy\ \D(y). 
\label{entropia_jamming}
\eeq
We expect that the entropy diverges as $\log{1/p} \simeq \log{\scl}$. This means that the term between square parentheses on the right hand side of (\ref{entropia_jamming}) must vanish. This gives a condition for the jamming point $\eta_J$ \cite{RUYZ15}.

\subsection{Asymptotes and scaling of $\wh{P}$ and $\wh{f}$}
In order to show that the scaling equations (\ref{scaling_eqs}) have the same critical exponents as the ones derived in \cite{CKPUZ14JSTAT} we need to show that the asymptotic behavior for $h\to \pm \infty$ of the initial conditions for $\wh f$ and $\wh P$ coincides with the one of \cite{CKPUZ14JSTAT}.
We start from the $\wh{f}$. 
Since the boundary condition for $\wh f$ is the same as the one in \cite{CKPUZ14JSTAT}, it trivially follows that also the asymptotic behavior is the same.
Indeed we have
\begin{eqnarray}
 \wh{f}(1/\scl,h\to-\infty) &=& -h^2/(2\gamma(1/\scl)),\\
 \wh{f}(1/\scl,h\to\infty) &=& 0,
\end{eqnarray}
and by inserting this asymptotes in the equation for $\wh{f}$ we get,
\begin{eqnarray}
\wh{f}(y,h\to-\infty) &=& -h^2/(2\gamma(y)),\\
\wh{f}(y,h\to\infty) &=& 0,
\end{eqnarray}
as in \cite{CKPUZ14JSTAT}. 
Conversely, the boundary condition for $\wh P$ is not the same as in \cite{CKPUZ14JSTAT}. However {one can easily see} that the asymptotic behavior is still the same.
Indeed, we have for $y=0$
\begin{eqnarray}
\wh{P}(0,h\to-\infty) &=& A(0)e^{B(0)h-D(0)h^2}\\
\wh{P}(0,h\to\infty) &=& e^{-\D(0)/2},
\end{eqnarray}
thanks to the fact that our $\wh{P}(0,h)$ is the convolution of a $\Theta$ function with a normalized Gaussian. 
We can again plug these asymptotes (and those of $\wh{f}$) in the equation for $\wh{P}$ in (\ref{scaling_eqs}), to get
\begin{eqnarray}
\wh{P}(0,h\to-\infty) &=& A(y)e^{B(y)h-D(y)h^2}\\
\wh{P}(0,h\to\infty) &=& e^{-\D(y)/2},
\end{eqnarray}
where the equations for $A,B$ and $D$ are the same as in \cite{CKPUZ14JSTAT}. 

We now look for a solution for $\wh{P}$ and $\wh{f}$ at large $y$. We conjecture that $\D(y)\simeq \Delta_\infty y^{-\kappa}$ for large $y$, which through the \eqref{eq:deltaofgamma} implies that $\gamma(y)\simeq \gamma_\infty y^{-c}$ with $c = \kappa-1$ and $\gamma_\infty = \frac{\kappa}{\kappa-1}\D_\infty$. We can then solve the equations for $A,B$ and $D$ for large $y$, and we get for $h\to-\infty$
\beq
\wh{P}(y,h) = A_\infty y^c e^{B_\infty h^cy^c-D_\infty h^2y^{2c}} = y^cp_0(hy^c).
\eeq
We can thus conjecture for $\wh{P}$ the same exact scaling that was used in \cite{CKPUZ14JSTAT}:
\beq
\wh{P}(y,h)\simeq\begin{cases}
             y^c p_0(hy^c)  & h\simeq-y^{-c}\\
             y^ap_1(hy^b) & |h|\simeq y^{-b}\\
             p_2(h) & h \gg y^{-b}\:.
            \end{cases}
\eeq
This scaling in turn requires that the function $p_1(z)$ must obey the boundary conditions
\beq
p_1(z)=\begin{cases}
             z^\theta & z\to\infty \\
             z^{-\alpha} & z\to - \infty
            \end{cases}
\eeq
where $\theta\equiv\frac{c-a}{b-c}$ and $\alpha = \frac{a}{b}$, as in \cite{CKPUZ14JSTAT} and \cite{perceptron}.

For what concerns the $\wh{f}$, we define as in \cite{CKPUZ14JSTAT} a function
\beq
\wh{j}(y,h) \equiv \wh{f}(y,h) + \frac{h^2\theta(-h)}{2\gamma(y)}. 
\eeq
Using the equation for $\wh{f}$ it is easy to see that, for all $y$
\begin{eqnarray}
\wh{j}(y,h\to-\infty) &=& \int_y^\infty \frac{du}{2u} \frac{\dot{\gamma}(u)}{\gamma(u)},\\
\wh{j}(y,h\to \infty) &=& 0.
\end{eqnarray}
For large $y$, again $\gamma(y) \simeq \gamma_\infty y^{-c}$, which means $\wh{j}(y,h\to-\infty) \simeq -c/(2y)$. So we can again conjecture the scaling form
\beq
\wh{j}(y,h) = -\frac{c}{2y}J(hy^b/\sqrt{\gamma_\infty}).
\eeq
with the boundary conditions $J(-\infty) = 1$ and $J(\infty)=0$. 

Now that we have the boundary conditions for the functions $J$ and $p_1$, all that we have to do is to plug them into the equations for $\wh{P}$ and $\wh{f}$ in order to get the equations for $p_1$ and $J$: since the scaling equations are the same as in \cite{CKPUZ14JSTAT} we get the same equations for $p_1$ and $J$.
The final step to show that the critical exponents here are the same as in \cite{CKPUZ14JSTAT} is to show that the \emph{marginal stability} equation is the same \cite{CKPUZ14JSTAT}.
{In the state following case, the equation for the marginal stability of fullRSB states is Eq. \ref{MG_SF} and it coincides with what has been obtained in \cite{CKPUZ14JSTAT}.}
We conclude that the scaling behavior of the solution of (\ref{scaling_eqs}) is the same as the one found in \cite{CKPUZ14JSTAT}, thus proving that the critical exponents
$a, b$ and $c$ and $\k$, $\th$ and $\a$ are the same as in \cite{CKPUZ14JSTAT}.

\section{Perturbative solution around the Gardner point}
In this section we want to investigate two main aspects of the Gardner transition point where the fullRSB solution appears.
On the one hand, we expect that at the Gardner transition the equilibrium dynamics inside a glassy state slow down and develop a power law {divergence of the relaxation time, i.e. \emph{critical slowing down}. We are thus interested in computing the associated dynamical exponent.\\ On the other hand, since the presence of the fullRSB hierarchy corresponds to an infinite set of timescales, we expect also that the elastic response of marginal glasses to be dependent on the timescale over which the system has been able to equilibrate.
This means that we can have a hierarchy of elastic moduli depending how much we have been able to equilibrate the system along the hierarchical structure of states. Thus in the following we compute the behavior of the elastic moduli close to the Gardner point.

\subsection{Computation of the $\l$ parameter}
Suppose to initialize the dynamics of a system of Hard Spheres inside a precise metastable state \cite{BBM96}. This can for example be done numerically using a planting procedure \cite{KZ09} as described in \cite{CJPRSZ15,MPR15}.
We want to investigate how the dynamics relaxes towards equilibrium within the metastable state.
In the stable glass phase the metastable state is ergodic and an exponential relaxation is {consequently observed}.
However as soon as the Gardner point is approached, we expect a dynamical slow down due to the appearance of an internal structure of substates within the metastable basin. 
Indeed, at this transition point, the relaxation becomes power law instead of exponential. 
Let us define a dynamical mean square displacement
\beq
\Delta_{D} (t)=\frac{d}{N}\sum_{i=1}^N|x_i(t)-x_i(0)|^2
\eeq
being $x_i(t)$ the position of the sphere $i$ at time $t$.
At the Gardner transition point we have
\beq
\Delta_D(t)\sim \Delta - A t^{-a}
\eeq
being $\Delta$ the solution of the equations (\ref{fullRSB_continue}) once a constant profile $\Delta(x)=\Delta$ is chosen; {the constant $A$ is expected to be positive.
We want to compute the exponent $a$. This is related to the so called \emph{exponent parameter} $\l$ \cite{PR12}
\beq
 \l = \frac{\Gamma(1-a)^2}{\Gamma(1-2a)}.
\eeq
It has been shown in \cite{PR12} that the exponent parameter can be computed from the replica approach. Indeed it is given by
\beq
\l=\frac{w_2}{w_1}
\eeq
and $w_1$ and $w_2$ are two cubic terms in the expansion of the free entropy around the RS solution at the Gardner point, defined in \cite{KPUZ13}. 
As proven in the SI of \cite{RUYZ15}, all the expressions for quadratic and cubic terms reported in \cite{KPUZ13} can be reused in the state following setting, just by redefining suitably the integral measure for computing averages
\beq
\la\mathcal O(\l)\ra_{RS} \equiv \int d\l\ \mathcal{O}(\l)\frac{e^{-\frac{(\l+\sqrt{\D})^2}{2}}}{\sqrt{2\pi}} \longrightarrow \la\mathcal O(\l)\ra_{SF} \equiv \int d\l\ \mathcal{O}(\l)G(\l), 
\eeq
where $G(\l)$ is defined in the SI of \cite{RUYZ15}. We can thus effortlessly write the expression for $\l$
\beq
\l = \frac{-8\wh\varphi_g w_2^{(I)}}{16/\D^3-8\wh\varphi_g w_1^{(I)}}
\eeq
where $w_1^{(I)}$ and $w_2^{(I)}$ are defined in \cite[Eq.(79)]{KPUZ13}. We can now eliminate the $\D^3$ factor, getting
\beq
\l = \frac{-8\wh\varphi_g \tilde{w}_2^{(I)}}{16-8\wh\varphi_g \tilde{w}_1^{(I)}}
\eeq
where $\tilde{w}_1^{(I)}$ and $\tilde{w}_2^{(I)}$ are defined as
\begin{eqnarray}
\tilde{w}_1^{(I)} &\equiv& -\la\Th_0(\l)^{s-1}\Gamma_1(\l,s)\ra,\\
\tilde{w}_2^{(I)} &\equiv& \frac{1}{2}\la\Th_0(\l)^{s-1}\Gamma_2(\l,s)\ra,
\end{eqnarray}
with
\beq
\begin{split}
\Gamma_2(\l,s)&=\left[ 2\left(\frac{\Th_1(\l)}{\Th_0(\l)}\right)^3 -3\frac{\Th_1(\l)\Th_2(\l)}{\Th_0^2(\l)}+\frac{\Th_3(\l)}{\Th_0(\l)}  \right]\left[  2\l^3+2(s-6)\left(\frac{\Th_1(\l)}{\Th_0(\l)}\right)^3  \right.  +\\
&\left.+ 3\frac{\Th_1(\l)}{\Th_0(\l)}\left[ 4\l\frac{\Th_1(\l)}{\Th_0(\l)}-(s-4)\frac{\Th_2(\l)}{\Th_0(\l)}\right] -6\l\left(\l\frac{\Th_1(\l)}{\Th_0(\l)}+\frac{\Th_2(\l)}{\Th_0(\l)}\right)+(s-2)\frac{\Th_3(\l)}{\Th_0(\l)}\right] \ , \\
\G_1(\l,s) &=\left[ 1+\frac{\Th^2_1(\l)}{\Th_0^2(\l)}-\frac{\Th_2(\l)}{\Th_0(\l)} \right]^2 \left[(s-3\l^2)+(s-6)\frac{\Th_1^2(\l)}{\Th_0^2(\l)}+6\l\frac{\Th_1(\l)}{\Th_0(\l)}-(s-3)\frac{\Th_2(\l)}{\Th_0(\l)}\right],
\end{split}
\eeq
and the $\Th_k(\l)$ functions are defined in \cite[Eq. (41)-(43)]{KPUZ13}.\\
We must evaluate the averages for $s\to 0$. This requires some caution as shown in section I.F.5 of the SI of \cite{RUYZ15} in the case of the replicon. In particular, we must pay attention to the asymptotic behavior of $\G_1$ and $\G_2$ for large $\l$. For $\G_2$ we have
\beq
\G_{2}(\l,s) \simeq \frac{2s}{\l^6} - \frac{12(4s-1)}{\l^8} + \dots,
\eeq
so the integral is convergent for every $s$ and no concern arises. For $\G_1$ we have, on the other hand,
\beq
\G_1(\l,s) \simeq s-\frac{3s}{\l^2} + \frac{21s-6}{\l^4} + \dots,
\eeq
which implies that
\beq
\la\Th_0(\l)^{s-1}\Gamma_1(\l,s)\ra = 1+\la\Th_0(\l)^{-1}\Gamma_1(\l,0)\ra,
\eeq
so we need to take into account the $+1$ correction for $s\to 0$. In summary, we have
\beq
\l = \frac{-4\wh\varphi_g A}{16+8\wh\varphi_g(1+B)},
\eeq
with
\begin{eqnarray}
A &=& \la\Th_0^{-1}(\l)\G_2(\l,0)\ra, \label{eq:defA}\\
B &=& \la\Th_0^{-1}(\l)\G_1(\l,0)\ra, \label{eq:defB}
\end{eqnarray}
where $A$ and $B$ must be computed numerically. 
We report the results of the numerical evaluation in Table \ref{table:lambdaMCT}.
\begin{table}[htb!]
\begin{tabular}{c@{\hspace{1cm}} c@{\hspace{1cm}}c}
\hline\hline
$\wh\varphi_g$ & $\l$ & $a$\\
\hline
4.8 & 0.702666 & 0.32402\\
4.9 & 0.560661 & 0.37718\\
5 & 0.509074 & 0.39267\\
5.25 & 0.437754 & 0.41210\\
5.5 & 0.393779 & 0.42313\\
5.87 & 0.351157 & 0.43319 \\
6 & 0.339808 & 0.43578\\
6.667 & 0.295692 & 0.44551\\
7 & 0.280148 & 0.44882\\
8 & 0.246892 & 0.45571\\
10.666 & 0.204280 & 0.46416\\
\hline
\end{tabular}
\caption{Our results for $\l$ and $1/a$ for various planting densities, including those studied in \cite{RUYZ15}\label{table:lambdaMCT}.}
\end{table}

\subsection{Perturbative 2RSB solution and shear moduli \label{subsec:shearmoduli}}
In this section we want to obtain a perturbative solution of the fullRSB equations just below the Gardner point where the system enters in the marginal glass phase.
In the fullRSB phase the profile $\Delta(x)$ has a continuous part with $\dot \Delta(x)<0$ and this survives up to the Gardner point.
However close to this transition, the continuum part is very small and we could produce a \emph{truncated} model that could describe the system just below the transition point as it has been done in appendix A of \cite{CFLPR11}.
We will not try this program here and we will only approximate the continuous profile of $\Delta(x)$ with a $2RSB$ ansatz that is expected to be a reasonable approximation.
The form of the mean square displacement profile {in this ansatz} will be a step function {defined as}
\beq
\Delta(x)=\begin{cases}
\Delta_1 &\mbox{if } x\in [0,\l]\\
\Delta_2 &\mbox{if } x\in\  ]\l,1]
\end{cases}\ \ \ \ \ \ \ \ \Delta_1>\Delta_2
\eeq
This corresponds to a dynamics in which the system first explores the configurations inside the innermost states of size $\Delta_2$ and then, on an infinite {(within mean-field)} timescale, relaxes to explore the entire metabasin of size $\Delta_1$.
Since the dynamics {takes place} on two timescales, we expect the response of the system to a small external perturbation to be as well characterized by two timescales. 
In particular we can consider the case wherin an infinitesimal shear strain is applied. {The glass will then respond linearly with a shear stress proportional to the strain, the proportionality constant being the \emph{shear modulus} or \emph{elastic modulus}.}
In \cite{YZ14} it has been shown that the shear modulus $\mu$ within a fullRSB ansatz is generally given by
\beq
\mu(x)=\frac{1}{\Delta(x)}
\eeq
so that at the 2RSB level we have two shear moduli
\beq
\begin{split}
\mu_1&=\frac{1}{\Delta_1}\\
\mu_2&=\frac{1}{\Delta_2}
\end{split}
\eeq
We want to show that when approaching the Gardner point we have
\beq
\frac{1}{\mu_1}-\frac{1}{\mu_2}=\Delta_1-\Delta_2\propto \wh \f-\wh \f_G
\eeq
In order to do this we want to start from the perturbative solution in the $\a_{ab}$ matrix that has already been studied in \cite{CKPUZ14JSTAT}. 
Since we know that the unstable mode is the replicon one, we can search for a perturbative solution that is {projected} along this {mode}. In this way we set the form of the perturbative 2RSB ansatz as
\beq
\a_{ab} \equiv \a_1(\d_{ab}s-1) + \d\a_2(1-\d_{ab})\left(\frac{s-s_1}{1-s_1}I_{ab}^{s_1}+I^s_{ab}-I^{s_1}_{ab}\right).
\label{eq:2RSBalpha}
\eeq
Recalling that $\D_{ab} = \a_{aa} + \a_{bb} -2 \a_{ab}$ we get
\begin{eqnarray}
\D_2 &=& 2\a_d - \a_{ab\in s_1}\\
\D_1 &=& 2\a_d - \a_{ab\notin s_1} 
\end{eqnarray}
where the notation $ab\in s_1$ means that the $ab$th matrix element is in one of the $s_1\times s_1$ sized blocks of the 2RSB matrix. Using the \eqref{eq:2RSBalpha} we get from this
\begin{eqnarray}
\D_2 &=& 2(s-1)\a_1 - 2\d\a_2\left(\frac{s-s_1}{1-s_1}\right) + 2\a_1,\\
\D_1 &=& 2(s-1)\a_1 - 2\d\a_2 + 2\a_1,
\end{eqnarray}
and in the limit $s\to0$ we get
\begin{eqnarray}
\D_2 &=& \D + 2\d\a_2\frac{s_1}{1-s_1},\\
\D_1 &=& \D-2\d\a_2,
\end{eqnarray}
where $\D$ is the MSD of the 1RSB solution and we have used $2s\a_1 = \D$ \cite{CKPUZ14JSTAT}. We now recall that
$$
\d\a_2 = -\frac{2\wh\l_R}{W}
$$
where $\wh\l_R$ is related to the replicon mode and $W$ is defined in \cite{CKPUZ14JSTAT}. We also recall that $\l_R>0$ and $0<s_1<1$ in the Gardner phase. So we get
\beq
\D_1-\D_2 = -2\d\a_2\frac{1}{1-s_1} = \frac{4\wh\l_R}{W}\frac{1}{1-s_1} \simeq C(\wh\varphi-\wh\varphi_G),
\eeq 
where the last equality is due to the fact that $\l_R$ is linear in $\wh\varphi-\wh\varphi_G$ near the Gardner point and $C$ is a negative constant.

\section{The $\chi_4$ susceptibility at the Gardner transition point}
The Gardner transition point can be detected \cite{CJPRSZ15} coming from the stable glass phase by looking at the so called $\chi_4$ susceptibility \cite{BBBCS11}.
Let us consider again the dynamical mean square displacement defined in the previous section.
We can look at its fluctuations that are given by {a dynamical susceptibility $\chi(t)$ \cite{BBBCS11}}
\beq
\chi(t)=\langle \Delta_D^2(t)\rangle-\langle\Delta_D(t)\rangle^2\:, 
\eeq
where the brackets are used to denote the average over the thermal history of the system.
The $\chi(t)$ is a dynamical quantity, but we can however focus on its large time behavior when $\langle\Delta_D(t)\rangle\to \Delta$.
In this case $\chi(t)\to \chi_4$, where $\chi_4$ is a static susceptibility (already well known in the context of spin glasses) that can be computed from a static approach \cite{FP00,FJPUZ12}.\\ 
To achieve this, we need to evaluate the Gaussian fluctuations around the replica symmetric saddle-point solution. These are defined in terms of a mass matrix that has been carefully defined in \cite{RUYZ15}.
We want to study the inverse of this quadratic operator {in order to obtain the value of the $\chi_4$ susceptibility}.
In \cite{RUYZ15}, the mass matrix has been derived in terms of the fluctuations of the elements of the matrix $\a_{ab}$ around the saddle point solution. However, since we want to evaluate the fluctuations of the mean square displacement matrix {we need to} compute the mass matrix that encodes for the Gaussian fluctuations {of the mean square displacement matrix $\D_{ab}$} around the saddle point solution.

This can be done in the following way. Let us consider two generic tensors $A_{ab;cd}$ and $B_{ab;cd}$ with the generic form
\beq
\begin{split}
A_{ab;cd} &= A_1T^1_{ab;cd} +A_2T^2_{ab;cd} + A_3T^3_{ab;cd}\\
B_{ab;cd} &= B_1T^1_{ab;cd} +B_2T^2_{ab;cd} + B_3T^3_{ab;cd}
\end{split}
\label{eq:genmassmatrix}
\eeq
where we have defined
\begin{eqnarray}
T^1_{ab;cd} &\equiv& \left(\frac{\d_{ac}\d_{bd}+\d_{ad}\d_{bc}}{2}\right)\\
T^2_{ab;cd} &\equiv& \left(\frac{\d_{ac} + \d_{bd}+\d_{ad}+\d_{bc}}{4}\right)\\
T^3_{ab;cd} &\equiv& 1.
\end{eqnarray}
We want to compute the tensor
$$
C_{ab;cd} = (A \otimes B)_{ab;cd} \equiv \sum_{e\neq f}^s A_{ab;ef}B_{ef;cd}.
$$
In order to do this, we must consider the products of the $T$ tensors. The $T^1$ tensor acts as a symmetrized identity, so
\begin{eqnarray}
 T^1\otimes T^1 &=& T^1,\\
 T^1\otimes T^2 &=& T^2,\\
 T^1\otimes T^3 &=& T^3.
\end{eqnarray}
Moreover we have
\beq
\begin{split}
T^2\otimes T^2=&\ \frac{(s-2)}{2} T^2 + \frac{1}{2}T^3,\\
T^2\otimes T^3 =&\ (s-1)T^3,\\
T^3\otimes T^3 =&\ s(s-1)T^3.
\end{split}
\eeq
So, the product of the two tensors $A$ and $B$ with the generic form  \eqref{eq:genmassmatrix} is a tensor $C$ with the same form,
$$
C_{ab;cd} = C_1T^1_{ab;cd} + C_2T^2_{ab;cd} + C_3T^3_{ab;cd}
$$
where the three coefficients entering in $C$ are expressed in terms of the $A$s and $B$s as
\begin{eqnarray}\label{finali}
C_1 &=& A_1B_1\nonumber\\
C_2 &=& \frac{s-2}{2}A_2B_2 + A_1B_2+A_2B_1 \label{eq:massmatrixproduct}\\
C_3 &=&\frac{A_2B_2}{2} + s(s-1)A_3B_3 + A_1B_3+A_3B_1 + (s-1)(A_2B_3+A_3B_2).\nonumber
\end{eqnarray}
If we set $C_1=1$, $C_2=C_3=0$, we get the equations for the coefficients of the inverse tensor $A^{-1}$:
\begin{eqnarray}
(A^{-1})_1 &=& \frac{1}{A_1}\nonumber\\
(A^{-1})_2 &=& -\frac{2A_2}{A_1(2A_1+(s-2)A_2)} \label{eq:inversemassmatrix}\\
(A^{-1})_3 &=&\frac{(A_2)^2-2A_1A_3+sA_2A_3}{A_1(2A_1+(s-2)A_2)(A_1+(s-1)(A_2+sA_3))}.\nonumber
\end{eqnarray}
Now that we have derived this formulae, let us consider the reduced mass matrix $M^{\D}_{ab;cd}$. Its definition is
\beq\label{Massa}
M^\D_{ab;cd} \equiv \frac{\d s[\wh\D]}{\d\D_{a<b}\d\D_{c<d}}
\eeq
where indices run from $m+1$ to $m+s$.
If we symmetrize this matrix
which means that we set
\beq
M_{a<b;c<d}^\D=M_{a>b;c<d}^\D=M_{a<b;c>d}^\D=M_{a>b;c>d}^\D
\eeq
we have that within the stable glass phase, before the Gardner point is reached, it has the form
\beq
M^\D_{ab;cd}=M_1^\D T^1+M_2^\Delta T^2+M_3^\D T^3\ \ \ \ \ \ a\neq b; c\neq d\:.
\eeq
In \cite{RUYZ15} we have computed 
\beq
M^\alpha_{ab;cd}\equiv \frac{\d s[\wh\D]}{\d\a_{a<b}\d\a_{c<d}}
\eeq
that {once symmetrized}, due to replica symmetry of the saddle point solution in the stable glass phase, has a similar structure
\beq
M^\D_{ab;cd}=M_1^\a T^1+M_2^\a T^2+M_3^\a T^3\ \ \ \ \ \ a\neq b; c\neq d\:.
\eeq
Since we know that $\hat \alpha$ and $\hat \D$ are related by $\D_{ab}=\a_{aa}+\a_{bb}-2\a_{ab}$ we can write a relation between the two mass matrices as
\beq
M^\a_{ab;cd} =\sum_{h<l}\sum_{m<n} \frac{\d\D_{hl}}{\d\a_{ab}}\frac{\d \D_{mn}}{\d\a_{cd}}M^\D_{mn;hl}.
\eeq
Now we need to compute $\frac{\d\D_{hl}}{\d\a_{ab}}$. This can be done in the following way.
Let us introduce an invertible tensor $U_{abcd}$ defined by that
\beq
\D_{ab} = \sum_{c\neq d}U_{abcd}\a_{cd}.
\eeq
and the tensor $U$ must be symmetric on the exchanges $a\leftrightarrow b$ and $c\leftrightarrow d$.
We have that
\beq
\frac{\d\D_{hl}}{\d\a_{a<b}}=2U_{hl;ab},
\eeq
{which implies}
\beq\label{direct}
M^\a_{a<b;c<d} =4\sum_{h<l}\sum_{m<n} U_{a<b,mn}U_{c<d,hl}M^\D_{mn;hl}=\sum_{h\neq l}\sum_{m\neq n} U_{a<b,mn}U_{c<d,hl}M^\D_{mn;hl}.
\eeq
To obtain $M^{\Delta}$ as a function of $M^\a$ we need to invert the operator $U$; this can be done in the following way. The change of coordinates between $\a_{ab}$ and $\D_{ab}$ can be written as
\beq
\begin{split}
\D_{ab} =\ &\a_{aa}+\a_{bb}-2\a_{ab},\\ 
=\ &-\sum_{c\neq a}^s\a_{ac} - \sum_{d\neq b}^s\a_{db} - 2\a_{ab},\\
=\ &(-2T^1_{ab;cd} -2T^2_{ab;cd})\a_{cd},
\end{split}
\eeq
Where we have used the fact that $\a$ is a Laplacian matrix and both $\D$ and $\a$ are symmetric. So we have that the tensor $U$ has the generic form \eqref{eq:genmassmatrix}, with coefficients $A_1=-2$, $A_2=-2$ and $A_3=0$. We can then use the equations \eqref{eq:inversemassmatrix} to get the coefficients of $U^{-1}$. The result is
\begin{eqnarray}
(U^{-1})_1 &=& -\frac{1}{2},\\
(U^{-1})_2 &=& \frac{1}{s},\\
(U^{-1})_3 &=& -\frac{1}{2s^2}.
\end{eqnarray}
Now that we have the expression of the tensor $U^{-1}$, we can obtain the mass matrix $M^{\D}$ from Eq. (\ref{direct}) in the following way
\beq
M^\D_{hl;mn}=\sum_{a\neq b}\sum_{c\neq d}[U^{-1}]_{hl;ab}[U^{-1}]_{mn;cd}M^\a_{ab;cd}\:.
\eeq
We are thus able the get the coeffients of the $M^\D$ mass matrix as a function of the coefficients of the $M^\a$. The result is
\begin{eqnarray}
M_1^\D &=& \frac{M_1^\a}{4}\\
M_2^\D &=& \frac{2M_2^\a-M_1^\a(s+2)}{2s^2}\\
M_3^\D &=& \frac{M_1^\a(s+3)-3M_2^\a+sM_3^\a}{4s^3}.
\end{eqnarray}

Now, we finally turn to the $\chi_4$ susceptibility. We want to compute
\beq\label{chidiv}
\chi_4=\la \D_{ab}^2 \ra - \la \D_{ab} \ra^2
\eeq
Let us define the full mass matrix that is given by
\beq\label{fullmass}
H_{ab;cd}=\frac{\delta^2 s}{\delta \Delta_{a<b}\delta \Delta_{c<d}}
\eeq
where $a,\ b,\ c,\ d$ run from $1$ to $m+s$.
The replicon mode is given by a variation of the mean square displacement matrix that is given by
\beq
\delta \Delta^R_{ab}: \sum_{b=m+1}^{m+s}\delta \Delta^R_{ab}=0
\eeq
and $\delta \Delta_{ab}=0$ if $a$ or $b$ are not in the group of slave replicas. 
The eigenvalue associated to the replicon is what we have computed in \cite{RUYZ15}.
Let us define 
\beq
\Omega_{ab;cd}=[H^{-1}]_{ab;cd}
\eeq
The fluctuation of the mean square displacement we want to compute is then given by
\beq
\la \D_{ab}^2 \ra - \la \D_{ab} \ra^2 = \Omega_{ab;ab}.
\eeq
Now $\Omega$ can be spectrally decomposed into two sectors: the first one is the replicon eigenspace that gives a singular contribution and the other one is in the space orthogonal to the replicon one. In other words we can write
\beq
\Omega_{ab;cd}=\frac{1}{\l_R^{\Delta}}P_{ab;cd}^{\parallel}+P_{ab;cd}^{\perp}
\eeq
where $\l_R^\D$ is the replicon eigenvalue of the mass matrix defined in Eq. (\ref{Massa}), $P^\parallel$ is the projector on the replicon subspace, and $P_{ab;cd}^\perp$ is not a projector but simply an operator that lives in the space orthogonal to the replicon one so that $P^\parallel \otimes P^{\perp}=0$. {With this decomposition we have}
\beq
\la \D_{ab}^2 \ra - \la \D_{ab} \ra^2 =\frac{1}{\l_R^{\Delta}}P_{ab;ab}^{\parallel} + P_{ab;ab}^{\perp},
\eeq
and if $P_{ab;ab}^{\perp}\neq 0$ as it can be generally expected, we notice that if we want to have both leading divergent contributions and finite corrections to the $\chi_4$, we need to invert the full mass matrix defined in Eq. (\ref{fullmass}), and not just the ``reduced'' one defined in the \eqref{Massa} and studied in \cite{RUYZ15}.\\
Let us first concentrate on the divergent contribution. 
We have {for the replicon}
\beq
\l_R^\D=\frac 12 M_1^\D
\eeq
This result is due to the fact that
\beq
\l_R^\D=\frac{\sum_{a<b;c<d}\delta \Delta_{ab}^RM_{ab;cd}^\D\delta \Delta_{cd}^R}{\sum_{c<d}\left(\delta \Delta_{cd}^R\right)^2}=\frac 12 \frac{\sum_{a\neq b;c\neq d}\delta \Delta_{ab}^RM_{ab;cd}^\D\delta \Delta_{cd}^R}{\sum_{c\neq d}\left(\delta \Delta_{cd}^R\right)^2}=\frac 12 M_1\:.
\eeq
Thus we have
\beq
\chi_4^\textrm{div}=\frac{8}{M_1^\alpha}P^\parallel_{ab;ab};
\eeq
in \cite{BR80} it has been shown that $P_{ab;ab}=3$ but in the appendix we give an alternative derivation of this result.
Finally we get
\beq
\chi_4^\textrm{div}=\frac{24}{M_1^{\alpha}}
\eeq
and this concludes the calculation of the most divergent part of the $\chi_4$ susceptibility.

\subsection{The finite corrections}
In the previous {paragraph} we have performed the calculation of the inverse of the reduced mass matrix $M^\D_{ab;cd}$. We now focus on the inverse of the full mass matrix for the theory $H_{ab;cd}$. We restrict ourselves to the case $m=1$ so that we have just one master replica.
The mass matrix $H_{(ab)^s;(cd)^s}$ in the $s$ sector {will be equal to the} reduced mass matrix discussed above. We now have to consider the part of the $M^\D$ tensor that stems from the fluctuations of the mean square displacement matrix elements $\D_{ab}$ that involve the $s$ replicas and the single master replica. This part of the tensor will produce the non-singular corrections to the $\chi_4$ susceptibilities\footnote{Note that non-singular correction arise not only from the mixing displacements between $s$-replicas and the master one but also from non singular terms of the inverse of the reduced mass matrix (\ref{Massa})}.
Due to replica symmetry, in order to specify the complete $H$ tensor we need two new sub-tensors $H_{1b;1d}$ and $H_{1b;cd}$, where all indices denoted with letters run from $1$ to $s+1$ unless otherwise specified. The most general form of the two tensors {compatible with replica symmetry} is
\begin{eqnarray}
H_{1b;1d} &=& A_1\d_{bd} + A_2\\
H_{1b;cd} &=& B_1\left(\frac{\d_{bc}+\d_{bd}}{2}\right) + B_2.
\end{eqnarray}
We now have to find the inverse tensor $\tilde{G} = 4G$ such that
$$
\tilde{G} \times H = \mathbb{1} =  \frac{1}{2} \tilde{G}\otimes H = 2T_1 \Longrightarrow G\otimes H = T_1
$$
where the $\otimes$ product and the tensor $T^1$ have exactly the same form as before with the only difference that their indices run from $1$ to $s+1$. The $\times$ product is the symmetrization of the $\otimes$ product, and the identity tensor is defined as $\mathbb{1} \equiv 2T_1$ because
\beq
\frac{\sum_{a<b;c<d}\delta \Delta_{ab}^RT_{ab;cd}^1\delta \Delta_{cd}^R}{\sum_{c<d}\left(\delta \Delta_{cd}^R\right)^2}=\frac 12 \frac{\sum_{a\neq b;c\neq d}\delta \Delta_{ab}^RT_{ab;cd}^1\delta \Delta_{cd}^R}{\sum_{c\neq d}\left(\delta \Delta_{cd}^R\right)^2}=\frac 12\:.
\eeq
In order to perform the inverse of the full mass matrix, let us look closer at the $\otimes$ product (let us fix $H_{ab;cd} = 0$ when $a=b$ or $c=d$)
\beq
(G\otimes H)_{ab;cd} = \sum_{e\neq f}G_{ab;ef}H_{ef;cd} = \sum_e \sum_f [G_{ae}]_{bf}[H_{ec}]_{fd} = \sum_e [\mathbf{G}_{ae}\cdot\mathbf{H}_{ec}]_{bd} = \left[\sum_e \mathbf{G}_{ae}\cdot\mathbf{H}_{ec}\right]_{bd}
\eeq
where $\mathbf{G}_{ae}$ and $\mathbf{H}_{ae}$ are matrices. This means that each four-index tensor $T_{ab;cd}$ can be seen as a two-index tensor (a matrix), where each coefficient identified by a couple of the 1st and 3rd indices is itself a matrix of reals whose coefficients are identified by the 2nd and 4th indices:
\beq
H_{ab;cd} = [H_{ac}]_{bd}.
\eeq
The $\otimes$ product can then be seen simply as a matrix product, with the difference that the coefficients of the ``matrices'' are matrices themselves whose product is implemented as a matrix product, instead of the usual product between reals.\\
We can then write our complete tensor $H$ as a block matrix
\beq
\begin{pmatrix}
\mathbf{H}_{11} & \mathbf{H}_{1b}\\
\mathbf{H}_{b1} & \mathbf{H}_{ab}^s
\end{pmatrix}
\label{eq:Masmatrix}
\eeq
Where each $\mathbf{H}_{ab}$ object is a matrix. Thus we have a block matrix with a single ``coefficient'' $\mathbf{H}_{11}$, a ``row'' $\mathbf{H}_{1b}$ and a $s\times s$ ``matrix'' $H^s$ that corresponds to our reduced mass tensor \eqref{Massa}. Since the product between tensors is implemented as a matrix product, we can just use the formulas for the inverse of a block matrix, keeping in mind that the ``coefficients'' are themselves matrices. The inverse of the matrix \eqref{eq:Masmatrix} in the $s$ block is then given by
\beq
(\mathbf{H}_{ab}^s - \mathbf{H}_{a1}\cdot\mathbf{H}^{-1}_{11}\cdot\mathbf{H}_{1b})^{-1}.
\eeq
We are interested in the second term, which contains the finite corrections. We need to compute
\beq
[\mathbf{H}_{a1}\cdot\mathbf{H}^{-1}_{11}\cdot\mathbf{H}_{1c}]_{bd} = \sum_e\sum_f [H_{a1}]_{be}[H^{-1}_{11}]_{ef}[H_{1c}]_{fd} = \sum_e\sum_f H_{ab;1e}H^{-1}_{1e;1f}H_{1f;cd}.
\label{eq:1stproduct}
\eeq
As we have said, the $\mathbf{H}_{11}$ is a matrix so we have to consider its inverse in the matrix sense. However, since
\beq
[H_{11}]_{ab} = A_1\d_{ab} + A_2
\eeq
so $\mathbf{H}_{11}$ is just an RS matrix and we already know that its inverse has the form
\beq
[H^{-1}_{11}]_{ab} = H^{-1}_{1a;1b} =  \Omega_1\d_{ab} + \Omega_2
\eeq
where the expressions of the $\Omega_i$ are
\begin{eqnarray}
 \Omega_1 &=& \frac{1}{A_1},\\
 \Omega_2 &=& -\frac{A_2}{A_1(A_1+sA_2)}.
\end{eqnarray}
We also know that
\beq
H_{ab;1e} = B_1\left(\frac{\d_{ae}+\d_{be}}{2}\right) +B_2.
\eeq
We can know compute the product \eqref{eq:1stproduct}. The calculation is standard so we skip right to the final result:
\beq
[\mathbf{H}_{a1}\cdot\mathbf{H}^{-1}_{11}\cdot\mathbf{H}_{1c}]_{bd} = B_1W_1\left(\frac{\d_{ac}+\d_{bd}+\d_{ad}+\d_{bc}}{4}\right) +(B_2W_1+B_1W_2+sB_2W_2)
\eeq
where
\begin{eqnarray}
W_1 &=& B_1\Omega_1\\
W_2 &=& B_1\Omega_2+B_2\Omega_1 + sB_2\Omega_2.
\end{eqnarray}
so we have a correction in the form 
\beq
C_2T^2 + C_3T^3,
\eeq
Which ensures that the tensor 
\beq
[\mathbf{H}_{ac}^s - \mathbf{H}_{a1}\cdot\mathbf{H}^{-1}_{11}\cdot\mathbf{H}_{1c}]_{bd}
\eeq
has the same general form written in the \eqref{eq:genmassmatrix}:
\beq
[\mathbf{H}_{ac}^s - \mathbf{H}_{a1}\cdot\mathbf{H}^{-1}_{11}\cdot\mathbf{H}_{1c}]_{bd} = M^\D_1T^1_{ab;cd} + (M^\D_2-B_1W_1)T^2_{ab;cd} + (M^\D_3-B_2W_1-B_1W_2 - sB_2W_2)T^1_{ab;cd},
\eeq
which means that we can perform the inverse and compute the fluctuations using the formulas \eqref{eq:inversemassmatrix}. We skip directly to the final result:
\beq
\chi_4 = 2\frac{3B_1^4+4A_1B_1B_2M^\D_1+B_1^2(3A_1(M_1^\D-2M_2^\D)-2A_2M_1^\D)+A_1^2((M_1^\D)^2-3M_1^\D M_2^\D +3(M_2^\D)^2-2M_1^\D M_3^\D)}{M_1^\D(B_1^2 + A_1(M_1^\D-M_2^\D))^2}
\eeq

\section{Conclusions and perspectives}
In this work we have shown how to follow glassy states under compression or shear strain in the regime where the replica symmetry is {continuously} broken.
Beyond the Gardner transition point, the system is described by a fullRSB solution, and we have shown how to obtain the fullRSB equations in the marginal glass phase.
Using them, we have been able to show that {glassy states can be followed all the way to the jamming point, in whose vicinity} the fullRSB solution develops a scaling regime characterized by the same critical exponents that have been obtained in \cite{CKPUZ14NatComms}. Moreover, we have computed the dynamical critical exponent of the relaxation dynamics close to the Gardner point, and we have shown that the fluctuations of the mean square displacement diverge at the Gardner transition by computing the most singular contribution to the $\chi_4$ susceptibility.

Despite the fact that the numerical solution of the equations is simple enough in a compression protocol, the same is not true for the evolution of glassy states under a shear strain, wherein the numerical code for the resolution of the fullRSB equations shows instability problems. In this case the numerical and theoretical analysis close to the yielding transition may require a more refined code and may involve again a scaling-type analysis like the one performed at the jamming point. {This problem is however both technical and conceptual, and not of immediate solution: the properties of the yielding transition are right now object of  intense study and debate, and the presence of an underlying criticality (with an associated set of critical exponents) is still a very much open issue \cite{KLP10, LGRW15, HJPS15, LLRW14}.} For these reasons we leave for a future work the detailed study of the mean field predictions for the yielding transition.\\
{Since the picture for the yielding transition that comes from \cite{RUYZ15} and the present work is that of a spinodal point, a first step is study the stability of the glassy minimum as the transition is approached. The critical mode that is relevant in such a case is the so-called \emph{longitudinal} mode of the hessian matrix defined in \cite{RUYZ15}, generalized in this case to the fullRSB ansatz. A computation of this mode should provide new insights in the physics of yielding.}

{Another natural continuation of this work is the study of Athemal Quasi Static (AQS) shear protocols \cite{ML06}; the typical stress-strain curves measured in these situations can be tought of as made of a smooth, ``averaged'' part, plus a highly intermittent part characterized by sharp stress drops, or avalanches \cite{CR00}, a phenomenon analogous to Barkhausen noise in spin glasses \cite{LMW10,LMW12}. It is interesting to note that the slope of the smooth part and of the segments in the fluctuating part can be respectively linked to two different shear moduli, in agreement with the discussion of paragraph \ref{subsec:shearmoduli}.
\\
The small jumps that characterize the response of the system (both for strained jammed packings \cite{CR00} and spin glasses \cite{LMW10}) are known to be power-law distributed for small amplitudes. The power-law exponent has been computed from first principles in the case of spin glasses \cite{LMW12}, and a calculation of the exponent for Hard Spheres can in principle be accomplished by generalizing the computation of \cite{LMW12}. In both cases, the fullRSB equations we derived in this work will be a necessary starting point.}

\section{Acknowledgments}
We would like to warmly thank G. Parisi and F. Zamponi for giving us the opportunity to use and adapt the numerical code to solve the fullRSB equations.
Moreover we would like to thank G. Biroli, S. Franz, J. Kurchan, T. Maimbourg, G. Parisi, H. Yoshino and F. Zamponi for illuminating discussions.
The research leading to these results has received funding from the European Research Council under the European Union's Seventh Framework Programme (FP7/2007-2013) / ERC grant agreement n° [247328]. P.U. Acknowledges the support of the ERC grant NPRGGLASS.

\appendix
\section{Calculation of the replicon projector}
%
{
We construct the operator that project on the replicon subspace.
This projector is of the form
\beq
P_{ab;cd}^\parallel=P_1\left(\frac{\delta_{ac}\delta_{bd}+\delta_{ad}\delta_{bc}}{2}\right)+P_2\left(\frac{\delta_{ac}+\delta_{bd}+\delta_{ad}+\delta_{bc}}{4}\right)+P_3
\eeq
and the replica indexes $a,\ b,\ c,\ d$ must be all in the interval $[m+1,m+s]$ because otherwise we fix the matrix element of $P^\parallel$ to zero.
The projector must satisfy the following tree properties
\beq
\begin{split}
&\delta \Delta_{cd}^R=\sum_{c\neq d}P_{ab;cd}^\parallel\delta \Delta_{cd}^R\\
&P^\parallel_{ab;ef}=\sum_{c\neq d}P_{ab;cd}^\parallel P_{cd;ef}^\parallel\\
&0=\sum_{c\neq d}P_{ab;cd}^\parallel
\end{split}
\eeq
The first equation tells that the projector on the replicon subspace leaves unchanged a vector in the replicon subspace. The second equation is the normalization of the projector and the last equation tells that the longitudinal eigenspace is in the kernel of the projector.
The first equation gives directly $P_1=1$.
The second equation instead gives that
\beq
P_2=\frac{2}{2-s}
\eeq
and the last relation instead gives
\beq
P_3=\frac{1}{(s-1)(s-2)}
\eeq
It can be easily verified that the anomalous eigenspace is in the kernel of the projector. Indeed, an anomalous eigenvector is a vector of the form
\beq
\delta_A\Delta_{ab}=\frac{1}{2}\left(\phi_a+\phi_b\right)\ \ \ \ \ \ \ \ \ \sum_{a=m+1}^{m+s}\phi_a=0
\eeq
and we find that 
\beq
\sum_{cd}P^\parallel_{ab;cd}\delta_A\Delta_{cd}=0
\eeq
}


\bibliographystyle{mioaps}
\bibliography{HS}

\end{document}